\documentclass[12pt]{article}
\usepackage{amssymb}

\usepackage{amsmath}
\usepackage[unicode,bookmarks,bookmarksopen,bookmarksopenlevel=2,colorlinks,linkcolor=blue,citecolor=green]{hyperref}

\def\comment#1{}
\input{tcilatex}

\begin{document}

\title{Algebro-geometric approach in the theory of integrable hydrodynamic
type systems}
\author{Maxim V. Pavlov}
\date{}
\maketitle

\begin{abstract}
The algebro-geometric approach for integrability of semi-Hamiltonian
hydrodynamic type systems is presented. This method is significantly
simplified for so-called \textit{symmetric} hydrodynamic type systems.
Plenty interesting and physically motivated examples are investigated.
\end{abstract}

\tableofcontents

\textit{keywords}: Riemann surface, chromatography, hydrodynamic type
system, generalized hodograph method.

MSC: 35L40, 35L65, 37K10;\qquad PACS: 02.30.J, 11.10.E.

\section{Introduction}

The theory of the integrable hydrodynamic type systems%
\begin{equation}
u_{t}^{i}=\upsilon _{j}^{i}(\mathbf{u})u_{x}^{j}\text{, \ \ \ \ \ \ }%
i,j=1,2,...,N  \label{0}
\end{equation}%
was established by S.P. Novikov, B.A. Dubrovin (see \textbf{\cite{Dubr+Nov}}%
) and S.P. Tsarev (see \textbf{\cite{Tsar}}). This differential-geometric
approach has been developed also by E.V. Ferapontov (see, for instance, 
\textbf{\cite{Fer+trans}}, \textbf{\cite{Fer+Tsar} }and other references
therein), I.M. Krichever (see, for instance, \textbf{\cite{Krich+kp}} and
other references therein), O.I. Mokhov (see, for instance, \textbf{\cite%
{Mokh}} and other references therein) and by the author (see, for instance, 
\textbf{\cite{Maks+Tsar}} and other references therein). Also, B.A. Dubrovin
and I.M. Krichever (see, for instance, \textbf{\cite{Dubr}}, \textbf{\cite%
{Krich}} and other references therein), Yu. Kodama and J. Gibbons (see, for
instance, \textbf{\cite{Gib+Yu}}, \textbf{\cite{Kod+water} }and other
references therein) used algebro-geometric approach in the theory of
integrable hydrodynamic type systems, which are dispersionless limits of
integrable dispersive equations (or in most general case the Whitham
equations, i.e. obtained by the Whitham averaging method of multi-phase
solutions of dispersive systems). Thus, all corresponding information like
the Riemann surfaces, a quasi-momentum and a quasi-energy can be
reconstructed. Then (as usual in the algebro-geometric approach) generating
functions of conservation laws and commuting flows can be found
automatically.

This paper is devoted to the algebro-geometric approach for hydrodynamic
type systems, whose origin is \textbf{unknown}. For simplicity in this paper
we restrict our consideration on \textit{symmetric} hydrodynamic type
systems, because just in this case a generating function of conservation
laws in fact is given in advance. In all other cases derivation of a
generating function of conservation laws is separate and complicated
computational problem, which will be investigated in details in another
publication devoted integrable hydrodynamic chains. The next step is a
computation of the equation of the Riemann surface. The corresponding linear
ODE system can be solved if it is invariant under a some Lie group symmetry.
For instance, if the symmetric hydrodynamic type system is homogeneous (this
is a typical formulation of physically motivated examples), then the
corresponding generating function of conservation laws and the equation of
the Riemann surface are homogeneous too.

Moreover, the generalized hodograph method established by S.P. Tsarev in 
\textbf{\cite{Tsar}} is based on a concept of the Riemann invariants. In
this paper we suggest an alternative approach based on a conservative form
of hydrodynamic type systems. Most physically motivated problems are given
in such form.

The paper is organized in the following order. In the second section a
semi-Hamiltonian (integrability) property for hydrodynamic type systems is
reformulated for a conservative form. The method allowing immediately to
construct generating function of conservation laws and the corresponding
Riemann surface is established. In the third section we briefly describe
Tsarev's observations improving plenty calculations. In the fourth section
the chromatography system as the most interesting and very complicated
example of symmetric hydrodynamic type systems is investigated. In the fifth
section the generalized hodograph method adopted to a conservative form is
considered in details. Several different sub-classes of hydrodynamic type
systems are presented. In the sixth section homogeneous hydrodynamic type
systems are considered. In such case a computation of the Riemann surface
can be found in quadratures. In the seventh section integrable hydrodynamic
chains as a natural generalization of the symmetric hydrodynamic type
systems are discussed. In the eighth section the Hamiltonian chromatography
hydrodynamic type system is considered. Corresponding hydrodynamic chain is
found as well its local Hamiltonian structure. In the ninth section the
linearly degenerate case is investigated. In the conclusion the main open
problem is considered.

\section{\textit{Symmetric} hydrodynamic type systems}

The hydrodynamic type system (\textbf{\ref{0}}) written in the Riemann
invariants%
\begin{equation}
r_{t}^{i}=\mu ^{i}(\mathbf{r})r_{x}^{i}\text{, \ \ \ \ \ }i=1,2,...,N
\label{ri}
\end{equation}%
is integrable by the generalized hodograph method (see \textbf{\cite{Tsar}})
iff the semi-Hamiltonian condition%
\begin{equation}
\partial _{j}\frac{\partial _{k}\mu ^{i}}{\mu ^{k}-\mu ^{i}}=\partial _{k}%
\frac{\partial _{j}\mu ^{i}}{\mu ^{j}-\mu ^{i}}\text{, \ \ \ \ \ \ \ }i\neq
j\neq k  \label{semi}
\end{equation}%
is valid identically (here $\partial _{k}\equiv \partial /\partial r^{k}$;
the semi-Hamiltonian condition also can be written in arbitrary field
variables $u^{k}$, see \textbf{\cite{SS}}). Then the hydrodynamic type
system (\textbf{\ref{0}}) has infinitely many conservation laws and
commuting flows%
\begin{equation}
r_{y}^{i}=w^{i}(\mathbf{r})r_{x}^{i},  \label{com}
\end{equation}%
parameterized by $N$ arbitrary functions of a single variable.

\textbf{Remark}: Characteristic velocities $w^{i}(\mathbf{r})$ satisfy the
linear PDE system (see \textbf{\cite{Tsar}})%
\begin{equation}
\partial _{k}w^{i}=\frac{\partial _{k}\mu ^{i}}{\mu ^{k}-\mu ^{i}}%
(w^{k}-w^{i})\text{, \ \ \ \ \ \ \ }i\neq k,  \label{a}
\end{equation}%
which cannot be solved explicitly in general case, because this is a linear
system with variable coefficients. However, in this paper we are able to
avoid this problem, because the generalized hodograph method can be
formulated via arbitrary field variables (see the beginning of the section 
\textbf{6}). The algebro-geometric approach is based on the concept of the
Riemann surface, where all conservation laws and commuting flows can be
found (see below) via the Riemann invariants as well as conservation law
densities $u^{k}$, which appear in this framework in a natural way. The
system (\textbf{\ref{a}}) has the general solution parameterized by $N$
arbitrary functions of a single variable. The \textit{generalized hodograph
method} established by S.P. Tsarev (see \textbf{\cite{Tsar}}) leads to the
general solution written in implicit form as an algebraic system%
\begin{equation}
x+\mu ^{i}t=w^{i}\text{, \ \ \ \ \ \ \ \ }i=1,2,...,N  \label{alg}
\end{equation}%
for the semi-Hamiltonian hydrodynamic type system (\textbf{\ref{ri}}). In
this paper several tools useful for constructing of commuting flows are
suggested in the section \textbf{5}.

Plenty hydrodynamic type systems are known from physical applications, which
have the very special conservative form%
\begin{equation}
u_{t}^{i}=\partial _{x}\psi (u^{1},u^{2},...,u^{N};u^{i}),  \label{1}
\end{equation}%
where the sole function $\psi (u^{1},u^{2},...,u^{N};p)$ is invariant under
permutation of the first $N$ entries (also we restrict our consideration on
the case, when the function $\psi (\mathbf{u};p)$ is \textbf{nonlinear} with
respect to $p$, see details in the last section \textbf{9}). We call such
hydrodynamic type systems the \textit{symmetric} hydrodynamic type systems.
The theory presented below can be easily extended on more general symmetric
classes, for instance on%
\begin{eqnarray*}
u_{t}^{i} &=&\partial _{x}\psi (u^{1},u^{2},...,u^{N};\upsilon ^{1},\upsilon
^{2},...,\upsilon ^{M};u^{i})\text{,} \\
&& \\
\upsilon _{t}^{i} &=&\partial _{x}g(u^{1},u^{2},...,u^{N};\upsilon
^{1},\upsilon ^{2},...,\upsilon ^{M};\upsilon ^{i}),
\end{eqnarray*}%
where $N$ and $M$ are arbitrary integers. Moreover, we shall demonstrate
below (in a set of examples) that this is not necessary restriction.
However, in the symmetric case this theory is very effective.

Let us introduce the matrix $A_{i}^{k}$ given by%
\begin{equation}
A_{i}^{k}(\mathbf{u};p)=\left( \frac{\partial \psi }{\partial p}|_{p=u^{i}}-%
\frac{\partial \psi }{\partial p}\right) \delta _{i}^{k}+\frac{\partial \psi 
}{\partial u^{i}}|_{p=u^{k}}  \label{6}
\end{equation}%
and formulate the \textit{phenomenological} algebro-geometric approach.

\textbf{Statement 1}: \textit{If the symmetric hydrodynamic type system} (%
\textbf{\ref{1}}) \textit{is integrable, then this system has the generating
function of conservation laws}%
\begin{equation}
p_{t}=\partial _{x}\psi (u^{1},u^{2},...,u^{N};p).  \label{2}
\end{equation}

If the generating function of conservation laws (\textbf{\ref{2}}) is
consistent with the hydrodynamic type system (\textbf{\ref{1}}), then%
\begin{equation}
\frac{\partial p}{\partial u^{i}}=B_{i}^{k}\frac{\partial \psi }{\partial
u^{k}},  \label{9}
\end{equation}%
where the matrix $B_{i}^{k}(\mathbf{u};p)$ is an inverse matrix to the
matrix $A_{i}^{k}$ (see (\textbf{\ref{6}})). This is $N$ ODE's of the first
order for every \textit{fixed} index $i$, where any of them can be written
in the form $dy/dx=f(x,y)$.

Let us introduce the function $\lambda (u^{1},u^{2},...,u^{N};p)$ determined
by $N$ \textit{linear} PDE's of the first order (see (\textbf{\ref{9}}))%
\begin{equation}
A_{i}^{k}\frac{\partial \lambda }{\partial u^{k}}+\frac{\partial \psi }{%
\partial u^{i}}\frac{\partial \lambda }{\partial p}=0.  \label{7}
\end{equation}

\textbf{Definition 1}: \textit{The function} $\lambda (\mathbf{u};p)$ 
\textit{satisfying} (\textbf{\ref{7}}) \textit{is said to be the equation of
the Riemann surface}.

If the linear PDE system (\textbf{\ref{7}}) has an integration factor then
the equation of the Riemann surface can be found in quadratures%
\begin{equation*}
d\lambda =\frac{\partial \lambda }{\partial p}\left[ dp-B_{k}^{n}\frac{%
\partial \psi }{\partial u^{n}}du^{k}\right] .
\end{equation*}%
If, for instance, the hydrodynamic type system (\textbf{\ref{1}}) is \textit{%
homogeneous} (see examples below), then functions $\psi (\mathbf{u};p)$ and $%
\lambda (\mathbf{u};p)$ are homogeneous too. Then the integration factor can
be found by using the Euler theorem%
\begin{equation*}
\lambda =p\frac{\partial \lambda }{\partial p}+u^{k}\frac{\partial \lambda }{%
\partial u^{k}}.
\end{equation*}

\textbf{Statement 2}: \textit{A deformation of the Riemann surface
determined by the equation }$\lambda (\mathbf{u};p)$ \textit{satisfies the
Gibbons equation}%
\begin{equation}
\lambda _{t}-\frac{\partial \psi }{\partial p}\lambda _{x}=\frac{\partial
\lambda }{\partial p}[p_{t}-\partial _{x}\psi (\mathbf{u};p)].  \label{3}
\end{equation}

The Gibbons equation (first introduce in \textbf{\cite{Gibbons}}) has three
distinguish features:

\textbf{1}. if one fixes $\lambda =\limfunc{const}$ (free parameter), then
one obtains (\textbf{\ref{2}}),

\textbf{2}. if one fixes $p=\limfunc{const}$ (free parameter), then one
obtains the kinetic equation (a collisionless Vlasov equation) written in
so-called Lax form%
\begin{equation*}
\lambda _{t}=\{\lambda \text{, }\mathbf{\hat{H}}\}=\frac{\partial \lambda }{%
\partial x}\frac{\partial \mathbf{\hat{H}}}{\partial p}-\frac{\partial
\lambda }{\partial p}\frac{\partial \mathbf{\hat{H}}}{\partial x},
\end{equation*}%
where $\mathbf{\hat{H}}=\psi (\mathbf{u};p)$.

\textbf{3}. if one choose coordinates, which are the Riemann invariants $%
r^{i}$ ($i=1,2,...,N$) determined by the condition $\partial \lambda
/\partial p=0$ (see (\textbf{\ref{7}})), then the corresponding hydrodynamic
type system (\textbf{\ref{1}}) can be written in the diagonal form (\textbf{%
\ref{ri}}) 
\begin{equation}
r_{t}^{i}=\frac{\partial \psi }{\partial p}|_{p=p^{i}}r_{x}^{i}\text{, \ \ \
\ \ }i=1,2,...,N,  \label{rim}
\end{equation}%
where the corresponding values $p^{i}$ can be expressed via these Riemann
invariants $r^{k}$. In this algebro-geometric construction the Riemann
invariants are the branch points $r^{i}=\lambda |_{\partial \lambda
/\partial p=0}$ of the Riemann surface (exactly as it is in the Whitham
theory, see \textbf{\cite{Dubr}} and \textbf{\cite{Krich}}).

\textbf{Remark}:\ The characteristic velocities $\mu ^{k}$ of hydrodynamic
type system (\textbf{\ref{0}}) can be found from algebraic system%
\begin{equation}
\det |\upsilon _{k}^{i}(\mathbf{u})-\mu \delta _{k}^{i}|=0.  \label{8}
\end{equation}%
All of them must be \textbf{distinct} in agreement with Tsarev's assumptions
(see \textbf{\cite{Tsar}}). Thus, if the hydrodynamic type system (\textbf{%
\ref{1}}) is integrable by the generalized hodograph method, then the values 
$p^{i}$ are given by (see (\textbf{\ref{rim}}))%
\begin{equation}
\mu ^{i}(\mathbf{u})\equiv \frac{\partial \psi }{\partial p}|_{p=p^{i}},
\label{4}
\end{equation}%
where characteristic velocities are determined from (\textbf{\ref{8}})%
\begin{equation}
\det A_{i}^{k}(\mathbf{u};p)=0.  \label{5}
\end{equation}

Suppose the hydrodynamic type system (\textbf{\ref{1}}) is semi-Hamiltonian.
Then such system must have $N$ series of conservation laws, which can be
obtained by the substitution of the formal series%
\begin{equation}
p^{(k)}=u^{k}+\lambda \upsilon ^{k}(\mathbf{u})+\lambda ^{2}w^{k}(\mathbf{u}%
)+...  \label{zak}
\end{equation}%
in (\textbf{\ref{2}}). The compatibility conditions of the first $N$ extra
conservation laws%
\begin{equation*}
\partial _{t}\upsilon ^{i}(\mathbf{u})=\partial _{x}\left[ \upsilon ^{i}(%
\mathbf{u})\frac{\partial \psi }{\partial p}|_{p=u^{i}}\right]
\end{equation*}%
with the hydrodynamic type system (\textbf{\ref{1}}) are equivalent the
semi-Hamiltonian property.

\textbf{Main statement}: \textit{The symmetric hydrodynamic type system} (%
\textbf{\ref{1}}) \textit{is semi-Hamiltonian iff the compatibility condition%
} $\partial _{i}(\partial _{k}p)=\partial _{k}(\partial _{i}p)$ \textit{is
fulfilled}.

\textbf{Comment}: Computation of this compatibility condition can be made in
the coordinates $u^{k}$ (see (\textbf{\ref{1}})) or in the Riemann
invariants (see (\textbf{\ref{rim}}) and details in the Conclusion). In both
cases the nonlinear PDE system in involution coincides with integrability
criterion for hydrodynamic type systems following from existence of $N$
conservation laws and vanishing Haantjes tensor (see \textbf{\cite{Haan}}, 
\textbf{\cite{SS}}).

\section{Tsarev's observations}

Let us consider the dispersionless limit of the vector NLS (see \textbf{\cite%
{Zakh}})%
\begin{equation}
u_{t}^{i}=\partial _{x}\left( \frac{(u^{i})^{2}}{2}+\sum \eta ^{k}\right) 
\text{, \ \ \ \ \ \ }\eta _{t}^{i}=\partial _{x}(u^{i}\eta ^{i})\text{, \ \
\ \ \ }i=1,2,...,N.  \label{za}
\end{equation}%
The eigenvalue--eigenfunction problem (cf. (\textbf{\ref{5}})) is%
\begin{equation*}
\left| 
\begin{array}{cc}
(u^{i}-\mu )\delta _{ik} & 1 \\ 
\eta ^{i}\delta _{ik} & (u^{i}-\mu )\delta _{ik}%
\end{array}%
\right| \left| 
\begin{array}{c}
q^{i} \\ 
s^{i}%
\end{array}%
\right| =0,
\end{equation*}%
Thus,%
\begin{equation}
q^{i}=\frac{1}{\mu -u^{i}}\sum s^{k}\text{, \ \ \ \ \ \ \ \ }s^{i}=\frac{%
\eta ^{i}}{(\mu -u^{i})^{2}}\sum s^{k}.  \label{vec}
\end{equation}

The \textit{\textbf{first} Tsarev observation} \textbf{\cite{Tsar+obs}} is
that the sum of the last $N$ equations yields an expression determining
characteristic velocities $\mu ^{k}$ (see (\textbf{\ref{8}}) and (\textbf{%
\ref{5}})) via very compact formula%
\begin{equation*}
1=\sum \frac{\eta ^{n}}{(\mu -u^{n})^{2}}.
\end{equation*}

The \textit{\textbf{second} Tsarev observation} \textbf{\cite{Tsar+obs}} is
that the above expression can be integrated once with respect to $\mu $%
\begin{equation}
\lambda =\mu +\sum \frac{\eta ^{n}}{\mu -u^{n}}  \label{rima}
\end{equation}

\textbf{Comment}: This equation of the Riemann surface can be obtained
directly from the spectral problem for the vector NLS (see \textbf{\cite%
{Zakh}}). The characteristic velocities $\mu ^{k}$ are determined by the
condition $\partial \lambda /\partial \mu =0$, where the Riemann invariants $%
r^{i}=\lambda |_{\partial \lambda /\partial \mu =0}\equiv \lambda |_{\mu
=\mu ^{i}}$ are branch points of the Riemann surface (see \textbf{\cite%
{Gibbons}}). A \textit{deformation of the Riemann surface} is described by
the Gibbons equation (see \textbf{\cite{Gibbons}} again)%
\begin{equation}
\lambda _{t}-\mu \lambda _{x}=\frac{\partial \lambda }{\partial \mu }\left[
\mu _{t}-\partial _{x}\left( \frac{\mu ^{2}}{2}+\sum \eta ^{n}\right) \right]
,  \label{gibs}
\end{equation}%
which connects (\textbf{\ref{za}}) with (\textbf{\ref{rima}}).

The \textit{\textbf{third} Tsarev observation} \textbf{\cite{Tsar+obs}} is
that the \textbf{flat} diagonal metric of the hydrodynamic type system (%
\textbf{\ref{za}}) written in the Riemann invariants (\textbf{\ref{ri}})%
\begin{equation*}
r_{t}^{i}=\mu ^{i}(\mathbf{r})r_{x}^{i}\text{, \ \ \ \ \ \ \ }i=1,2,...,2N
\end{equation*}%
is given by (see also \textbf{\cite{Tsar}})%
\begin{equation}
g_{ii}=\underset{\lambda =r^{i}}{\limfunc{res}}\left( \frac{\partial \lambda 
}{\partial \mu }\right) ^{2}d\mu .  \label{res}
\end{equation}%
Indeed, since the Hamiltonian structure of the dispersionless limit of the
vector NLS is%
\begin{equation*}
u_{t}^{i}=\partial _{x}\frac{\partial \mathbf{h}}{\partial \eta ^{i}}\text{,
\ \ \ \ }\eta _{t}^{i}=\partial _{x}\frac{\partial \mathbf{h}}{\partial u^{i}%
},
\end{equation*}%
the diagonal metrics (in Riemann invariants)%
\begin{equation*}
g^{ii}=\underset{k=1}{\overset{N}{\sum }}\frac{\partial r^{i}}{\partial u^{k}%
}\frac{\partial r^{i}}{\partial \eta ^{k}}
\end{equation*}%
is given by%
\begin{equation*}
g^{ii}=2\underset{k=1}{\overset{N}{\sum }}\frac{\eta ^{k}}{(\mu
^{i}-u^{k})^{3}},
\end{equation*}%
where (see (\textbf{\ref{rima}}) and cf. (\textbf{\ref{vec}}))%
\begin{equation*}
\frac{\partial r^{i}}{\partial u^{k}}=\frac{\eta ^{k}}{(\mu ^{i}-u^{k})^{2}}%
\text{, \ \ \ \ }\frac{\partial r^{i}}{\partial \eta ^{k}}=\frac{1}{\mu
^{i}-u^{k}}.
\end{equation*}%
Thus, $g^{ii}=\partial ^{2}\lambda /\partial \mu ^{2}|_{\mu =\mu ^{i}}$ in
agreement with (\textbf{\ref{res}}).

\section{Generalized chromatography}

The chromatography process (see, for instance, \textbf{\cite{Fer+Tsar}}) is
described by the hydrodynamic type system%
\begin{equation}
u_{t}^{i}=\partial _{x}\frac{(u^{i})^{\alpha }}{[1+\sum \gamma
_{k}(u^{k})^{\beta }]^{\varepsilon }}\text{, \ \ \ \ \ }i=1,2,...,N
\label{hrom}
\end{equation}%
where $\alpha ,\beta ,\varepsilon $ and $\gamma _{i}$ are constants. All
results presented in this section generalize results from \textbf{\cite%
{Fer+Tsar}} (see formulas \textbf{1}, \textbf{3}, \textbf{4}, \textbf{5}).
This hydrodynamic type system has the obvious couple of conservation laws%
\begin{equation*}
\partial _{t}\left[ \sum \gamma _{k}(u^{k})^{\beta -\alpha +1}\right]
=(\beta -\alpha +1)\partial _{x}\left[ \frac{\alpha -\beta \varepsilon }{%
\beta (1-\varepsilon )}\Delta ^{1-\varepsilon }-\Delta ^{-\varepsilon }%
\right] ,
\end{equation*}%
\begin{equation*}
\partial _{t}\Delta ^{\frac{\varepsilon -\beta \varepsilon +\alpha }{\alpha }%
}=\partial _{x}\left[ \frac{\beta (\alpha +\varepsilon -\beta \varepsilon )}{%
\alpha +\beta -1}\Delta ^{\varepsilon \frac{1-\beta -\alpha }{\alpha }}\sum
\gamma _{n}(u^{n})^{\alpha +\beta -1}\right] ,
\end{equation*}%
where $\Delta =1+\sum \gamma _{k}(u^{k})^{\beta }$.

If $\alpha =\beta -1$, then the hydrodynamic type system (\textbf{\ref{hrom}}%
) is Hamiltonian (see \textbf{\cite{Dubr+Nov}}, \textbf{\cite{Maks+Tsar}})%
\begin{equation}
u_{t}^{i}=\frac{1}{\gamma _{i}}\partial _{x}\frac{\partial \mathbf{h}}{%
\partial u^{i}},  \label{ham}
\end{equation}%
where the momentum density is $\Sigma \gamma _{k}(u^{k})^{2}$ and the
Hamiltonian density is $\mathbf{h}=\Delta ^{1-\varepsilon }$.

Suppose the above hydrodynamic type system is integrable for some values of
constants $\alpha ,\beta ,\varepsilon $ and $\gamma _{i}$. Then the
generating function of conservation laws given by%
\begin{equation*}
p_{t}=\partial _{x}\frac{p^{\alpha }}{\Delta ^{\varepsilon }}
\end{equation*}%
\textit{should} exist. It is easy to check that the compatibility conditions 
$\partial _{i}(\partial _{k}p)=\partial _{k}(\partial _{i}p)$ are valid iff $%
\alpha =\beta \varepsilon $, where (see (\textbf{\ref{9}}))%
\begin{equation}
\frac{\partial p}{\partial u^{i}}=\frac{\gamma _{i}(u^{i})^{\beta
-1}p^{\alpha }}{(u^{i})^{\alpha -1}-p^{\alpha -1}}\left[ \sum \frac{\gamma
_{n}(u^{n})^{\alpha +\beta -1}}{(u^{n})^{\alpha -1}-p^{\alpha -1}}-\frac{%
\alpha \Delta }{\beta \varepsilon }\right] ^{-1}.  \label{comp}
\end{equation}%
All first derivatives (see (\textbf{\ref{7}}))%
\begin{equation*}
\frac{\partial \lambda }{\partial u^{i}}=\varphi (\mathbf{u},p)p^{\beta
\varepsilon }\frac{\gamma _{i}(u^{i})^{\beta -1}}{(u^{i})^{\beta \varepsilon
-1}-p^{\beta \varepsilon -1}}\text{, \ \ \ \ \ \ }\frac{\partial \lambda }{%
\partial p}=\varphi (\mathbf{u},p)\left[ 1-p^{\beta \varepsilon -1}\sum 
\frac{\gamma _{n}(u^{n})^{\beta }}{(u^{n})^{\beta \varepsilon -1}-p^{\beta
\varepsilon -1}}\right]
\end{equation*}%
are determined up to integration factor $\varphi (\mathbf{u},p)$, which is
not found yet. Then the equation of the Riemann surface $\lambda (\mathbf{u}%
,p)$ can be found in quadratures%
\begin{equation*}
d\lambda =\varphi (\mathbf{u},p)\left( dp+\frac{q^{\frac{\beta +1}{\beta
\varepsilon -1}}}{\beta \varepsilon -1}\sum \gamma _{n}\frac{(w^{n})^{\frac{%
\beta }{\beta \varepsilon -1}-1}dw^{n}}{w^{n}-1}\right) ,
\end{equation*}%
if the integration factor is $\varphi (p)=p^{-1-\beta }$ (here we use the
substitutions $p=q^{\frac{1}{\beta \varepsilon -1}}$ and $u^{n}=(w^{n}q)^{%
\frac{1}{\beta \varepsilon -1}}$). Then, the integrable hydrodynamic type
system (\textbf{\ref{hrom}})%
\begin{equation}
u_{t}^{i}=\partial _{x}\frac{(u^{i})^{\beta \varepsilon }}{[1+\sum \gamma
_{k}(u^{k})^{\beta }]^{\varepsilon }}\text{, \ \ \ \ \ }i=1,2,...,N
\label{ful}
\end{equation}%
is connected with the equation of the Riemann surface%
\begin{equation*}
\lambda =\frac{p^{-\beta }}{\beta }+\frac{1}{\beta \varepsilon -1}\sum
\gamma _{k}(w^{k})^{\delta }F(1,\delta ,\delta +1,w^{k}),
\end{equation*}%
where $_{2}F_{1}(a,b,c,z)$ is a hyper-geometric function and $\delta =\beta
/(\beta \varepsilon -1)$. If $\delta =m/n$, where $m$ and $n$ are integers,
then the equation of the Riemann surface $\lambda (\mathbf{u},p)$ can be
found in elementary functions.

\textbf{Remark}: If $\beta \rightarrow \infty $, then the hydrodynamic type
system (\textbf{\ref{ful}}) reduces to%
\begin{equation}
u_{t}^{i}=\partial _{x}\frac{e^{\varepsilon u^{i}}}{[1+\sum \gamma
_{k}e^{u^{k}}]^{\varepsilon }}\text{, \ \ \ \ \ }i=1,2,...,N.  \label{exp}
\end{equation}%
The corresponding equation of the Riemann surface is%
\begin{equation*}
\lambda =e^{-p}+\frac{1}{\varepsilon }\sum \gamma _{n}(w^{n})^{1/\varepsilon
}F(1,\frac{1}{\varepsilon },\frac{1}{\varepsilon }+1,w^{n}),
\end{equation*}%
where $p=\ln q$ and $u^{n}=\ln (w^{n}q)$. Moreover, if $\varepsilon =1$, the
above hydrodynamic system has the local Hamiltonian structure (\textbf{\ref%
{ham}}) with the Hamiltonian density $\mathbf{h}=\ln \Delta $. The above
equation of the Riemann surface reduces to%
\begin{equation}
\lambda =e^{-p}-\sum \gamma _{n}\ln (e^{u^{n}-p}-1).  \label{rexp}
\end{equation}

\textbf{Remark}: If $\varepsilon \rightarrow 0$, then the hydrodynamic type
system (\textbf{\ref{ful}}) reduces to%
\begin{equation*}
u_{t}^{i}=\partial _{x}\ln \frac{(u^{i})^{\beta }}{1+\sum (u^{k})^{\beta }}%
\text{, \ \ \ \ \ }i=1,2,...,N,
\end{equation*}%
where the constants $\gamma _{k}$ are removed by appropriate scaling of the
field variables $u^{k}$. The corresponding equation of the Riemann surface is%
\begin{equation*}
\lambda =\frac{p^{-\beta }}{\beta }-\sum (w^{n})^{\delta }F(1,-\beta
,1-\beta ,w^{n}),
\end{equation*}%
where $u^{k}=w^{k}p$. If $\beta =1$, the corresponding hydrodynamic type
system (\textbf{\ref{ful}}) has the local Hamiltonian structure (\textbf{\ref%
{ham}}) with the Hamiltonian density%
\begin{equation*}
\mathbf{h}=\sum u^{m}(\ln u^{m}-1)-(1+\sum u^{m})[\ln (1+\sum u^{n})-1].
\end{equation*}

\section{The generalized hodograph method}

If the hydrodynamic type system (\textbf{\ref{0}}) is semi-Hamiltonian, then
the general solution parameterized by $N$ arbitrary functions of a single
variable is given in an implicit form by the algebraic system (cf. (\textbf{%
\ref{alg}}); see \textbf{\cite{Tsar}})%
\begin{equation}
x\delta _{k}^{i}+t\upsilon _{k}^{i}(\mathbf{u})=w_{k}^{i}(\mathbf{u}),
\label{tot}
\end{equation}%
where $w_{k}^{i}(\mathbf{u})$ are characteristic velocities of an arbitrary
commuting flow.

However, in this paper above, we present the approach producing the
generating function of conservation laws only. Thus, we need to extend this
mechanism to produce the generating function of conservation laws and
commuting flows simultaneously. In this paper we restrict our consideration
on three sub-cases connected with the Egorov conjugate curvilinear
coordinate nets (see \textbf{\cite{Maks+Tsar}}), with the orthogonal
coordinate nets (see \textbf{\cite{Tsar}}) and with so-called ``mirrored''
conjugate nets (see below). The general case will be considered elsewhere.

Nevertheless, the first step is a description of $N$ series of conservation
laws (see (\textbf{\ref{zak}})). They can be obtained by expansion in the B%
\"{u}rmann--Lagrange series (see, for instance, \textbf{\cite{Lavr}}) at the
vicinity of each singular point.

\textbf{Theorem 1 \cite{Lavr}}: \textit{The analytic function}%
\begin{equation*}
y=y_{1}(x-x_{0})+y_{2}(x-x_{0})^{2}+y_{3}(x-x_{0})^{3}+...
\end{equation*}%
\textit{can be inverted} ($y(x)\rightarrow x(y)$) \textit{as the B\"{u}%
rmann--Lagrange series}%
\begin{equation*}
x=x_{0}+x_{1}y+x_{2}y^{2}+x_{3}y^{3}+...,
\end{equation*}%
\textit{whose coefficients are}%
\begin{equation}
x_{n}=\frac{1}{n!}\underset{x\rightarrow x_{0}}{\lim }\frac{d^{n-1}}{dx^{n-1}%
}\left( \frac{x-x_{0}}{y}\right) ^{n}\text{, \ \ \ \ \ }n=1,2,...
\label{ryad}
\end{equation}

For example, the so-called ``waterbag'' hydrodynamic type system (see 
\textbf{\cite{Gib+Yu}}, \textbf{\cite{Kod+water}})%
\begin{equation}
a_{t}^{i}=\partial _{x}\left( \frac{(a^{i})^{2}}{2}+\sum \varepsilon
_{k}a^{k}\right)  \label{water}
\end{equation}%
is connected with the equation of the Riemann surface%
\begin{equation}
\lambda =p-\sum \varepsilon _{k}\ln (p-a^{k}).  \label{voda}
\end{equation}%
The main (so-called ``Kruskal'') series of conservation law densities can be
obtain by substitution of the Taylor series%
\begin{equation}
p=\lambda -\frac{\mathbf{H}_{0}}{\lambda }-\frac{\mathbf{H}_{1}}{\lambda ^{2}%
}-\frac{\mathbf{H}_{2}}{\lambda ^{3}}-...  \label{chu}
\end{equation}%
into the above expression if $\Sigma \varepsilon _{k}=0$. If $\Sigma
\varepsilon _{k}\neq 0$, then at first, the above equation of the Riemann
surface must be replaced on%
\begin{equation}
\lambda -\sum \varepsilon _{k}\ln \lambda =p-\sum \varepsilon _{k}\ln
(p-a^{k}),  \label{wat}
\end{equation}%
because the Gibbons equation is invariant under scaling $\lambda \rightarrow 
\tilde{\lambda}(\lambda )$. In both cases $\mathbf{H}_{k}$ are polynomials
with respect to field variables $a^{n}$.

Also, the ``waterbag'' hydrodynamic type system has $N$ infinite series of
conservation laws. At first let us rewrite the equation of the Riemann
surface in the form%
\begin{equation*}
\lambda =(p-a^{i})e^{-p/\varepsilon _{i}}\underset{k\neq i}{\prod }%
(p-a^{k})^{\varepsilon _{k}/\varepsilon _{i}}
\end{equation*}%
for any fixed index $i$. Then the infinite series of conservation laws (%
\textbf{\ref{zak}})%
\begin{equation}
p^{(i)}=a^{i}+h_{1}^{(i)}(\mathbf{a})\lambda +h_{2}^{(i)}(\mathbf{a})\lambda
^{2}+h_{3}^{(i)}(\mathbf{a})\lambda ^{3}+...  \label{ser}
\end{equation}%
can be obtained with the aid of B\"{u}rmann--Lagrange series (see \textbf{%
\cite{Lavr}}), which coefficients are determined by (see (\textbf{\ref{ryad}}%
))%
\begin{equation*}
h_{n}^{(i)}=\frac{1}{n!}\frac{d^{n-1}}{d(a^{i})^{n-1}}\left(
e^{na^{i}/\varepsilon _{i}}\underset{k\neq i}{\prod }(a^{i}-a^{k})^{-n%
\varepsilon _{k}/\varepsilon _{i}}\right) \text{, \ \ \ \ \ }n=1,2,...
\end{equation*}%
Thus, the first conservation laws are%
\begin{equation*}
h_{1}^{(i)}=e^{a^{i}/\varepsilon _{i}}\underset{k\neq i}{\prod }%
(a^{i}-a^{k})^{-\varepsilon _{k}/\varepsilon _{i}}\text{, \ \ \ }h_{2}^{(i)}=%
\frac{e^{2a^{i}/\varepsilon _{i}}}{\varepsilon _{i}}\left( 1-\underset{n\neq
i}{\sum }\frac{\varepsilon _{n}}{a^{i}-a^{n}}\right) \underset{k\neq i}{%
\prod }(a^{i}-a^{k})^{-2\varepsilon _{k}/\varepsilon _{i}},...
\end{equation*}

\subsection{Quasi-symmetric form}

The dispersionless limit of the vector NLS (\textbf{\ref{za}}) is a
degeneration of the ``waterbag'' hydrodynamic type system (\textbf{\ref%
{water}}). A substitution of the expansion (see, for instance, \textbf{\cite%
{Bogdan}}) $\tilde{u}^{(k)}=u^{k}+\eta ^{k}/\varepsilon ^{k}+...$ in (%
\textbf{\ref{voda}})%
\begin{equation*}
\lambda =\mu -\underset{k=1}{\overset{N}{\sum }}\varepsilon _{k}\ln \frac{%
\mu -\tilde{u}^{k}}{\mu -u^{k}}
\end{equation*}%
yields (\textbf{\ref{rima}}) if $\varepsilon ^{k}\rightarrow \infty $.

Let us consider again the generating function of conservation laws (see (%
\textbf{\ref{gibs}}))%
\begin{equation}
\mu _{t}=\partial _{x}\left( \frac{\mu ^{2}}{2}+A^{0}\right) .  \label{con}
\end{equation}%
for the Benney hydrodynamic chain (see \textbf{\cite{Benney}})%
\begin{equation}
A_{t}^{k}=A_{x}^{k+1}+kA^{k-1}A_{x}^{0}\text{, \ \ \ \ \ }k=0,1,2,...
\label{bm}
\end{equation}%
and substitute $N$ Taylor series (\textbf{\ref{ser}})%
\begin{equation}
\mu ^{(i)}=a^{i}+\lambda b^{i}+\lambda ^{2}c^{i}+...  \label{taylor}
\end{equation}

\textbf{1}. Suppose the function $A^{0}$ depends on $N$ field variables $%
a^{k}$ only. Then corresponding hydrodynamic type system is%
\begin{equation*}
a_{t}^{i}=\partial _{x}\left( \frac{(a^{i})^{2}}{2}+A^{0}(\mathbf{a})\right)
.
\end{equation*}%
However, this hydrodynamic type system is integrable iff the function $A^{0}(%
\mathbf{a})$ satisfies some nonlinear PDE system%
\begin{equation*}
(a^{i}-a^{k})\partial _{ik}A^{0}=\partial _{k}A^{0}\partial _{i}\left( \sum
\partial _{n}A^{0}\right) -\partial _{i}A^{0}\partial _{k}\left( \sum
\partial _{n}A^{0}\right) \text{, \ \ }i\neq k,
\end{equation*}%
\begin{equation*}
(a^{i}-a^{k})\frac{\partial _{ik}A^{0}}{\partial _{i}A^{0}\partial _{k}A^{0}}%
+(a^{k}-a^{j})\frac{\partial _{jk}A^{0}}{\partial _{j}A^{0}\partial _{k}A^{0}%
}+(a^{j}-a^{i})\frac{\partial _{ij}A^{0}}{\partial _{i}A^{0}\partial
_{j}A^{0}}=0\text{, \ \ }i\neq j\neq k,
\end{equation*}%
which is a consequence of the compatibility conditions $\partial
_{i}(\partial _{k}\mu )=\partial _{k}(\partial _{i}\mu )$, where%
\begin{equation*}
\partial _{i}\mu =\frac{\partial _{i}A^{0}}{\mu -a^{i}}\left( \sum \frac{%
\partial _{n}A^{0}}{\mu -a^{n}}-1\right) ^{-1}.
\end{equation*}%
Several such particular choices are described above.

\textbf{2}. Suppose the function $A^{0}$ depends on $N$ field variables $%
a^{k}$ and $M$ field variables $b^{k}$ (where $M$ \textbf{must be not exceed}
$N$) only. Then corresponding hydrodynamic type system is%
\begin{eqnarray*}
a_{t}^{i} &=&\partial _{x}\left( \frac{(a^{i})^{2}}{2}+A^{0}(\mathbf{a,b}%
)\right) \text{, \ \ \ \ \ }i=1,2,...,N, \\
&& \\
b_{t}^{j} &=&\partial _{x}(a^{j}b^{j})\text{, \ \ \ \ \ \ }j=1,2,...,M.
\end{eqnarray*}%
Simplest such example is (\textbf{\ref{za}}). This procedure can be extended
on other auxiliary field variables from (\textbf{\ref{taylor}}). For
instance, the third such sub-case is%
\begin{eqnarray*}
a_{t}^{i} &=&\partial _{x}\left( \frac{(a^{i})^{2}}{2}+A^{0}(\mathbf{a,b,c}%
)\right) \text{, \ \ \ \ \ }i=1,2,...,N, \\
&& \\
b_{t}^{j} &=&\partial _{x}(a^{j}b^{j})\text{, \ \ \ \ \ \ }j=1,2,...,M, \\
&& \\
c_{t}^{k} &=&\partial _{x}\left[ a^{k}c^{k}+\frac{1}{2}(b^{k})^{2}\right] 
\text{, \ \ \ \ \ }k=1,2,...,K,
\end{eqnarray*}%
where $K\leqslant M\leqslant N$.

\textbf{3}. Suppose the function $A^{1}$ (see (\textbf{\ref{bm}})) is a
function of the field variables $a^{k}$ only. Then corresponding
hydrodynamic type system is%
\begin{equation}
A_{t}^{0}=\partial _{x}A^{1}(\mathbf{a})\text{, \ \ \ \ \ \ }%
a_{t}^{i}=\partial _{x}\left( \frac{(a^{i})^{2}}{2}+A^{0}\right) \text{, \ \
\ \ }i=1,2,...,N.  \label{third}
\end{equation}

Suppose the function $A^{2}$ is a function of the field variables $a^{k}$
only. Then corresponding hydrodynamic type system is%
\begin{equation*}
A_{t}^{0}=\partial _{x}A^{1}\text{, \ \ \ \ \ \ }A_{t}^{1}=\partial
_{x}\left( A^{2}(\mathbf{a})+\frac{(A^{0})^{2}}{2}\right) \text{,\ \ \ \ \ \ 
}a_{t}^{i}=\partial _{x}\left( \frac{(a^{i})^{2}}{2}+A^{0}\right) \text{, \
\ \ \ }i=1,2,...,N.
\end{equation*}%
Thus, in general case one can consider $M$ first moments $A^{k}$ and $K$
first \textit{sets} $a^{i_{1}}$, $b^{i_{2}}$, $c^{i_{3}}$, ... (where the
index $i_{k}$ run all values from $1$ up to $N_{k}$, $k=1,2,...,K$) as field
variables of corresponding hydrodynamic type systems. In such case just the
latest moment $A^{M}$ depends on sets $a^{i_{1}}$, $b^{i_{2}}$, $c^{i_{3}}$,
...

Nevertheless, the theory established above is still working for these
hydrodynamic type systems, because all of them have the same generating
function of conservation laws (\textbf{\ref{con}}).

\textbf{Example}: The dispersionless limit of the vector Yajima--Oikawa
system (see \textbf{\cite{Maks+Tsar}}) is the hydrodynamic type system (cf. (%
\textbf{\ref{za}}) and (\textbf{\ref{third}}))%
\begin{equation}
A_{t}^{0}=\partial _{x}\sum \eta ^{n}\text{, \ \ \ \ }u_{t}^{k}=\partial
_{x}\left( \frac{(u^{k})^{2}}{2}+A^{0}\right) \text{, \ \ \ }\eta
_{t}^{k}=\partial _{x}(u^{k}\eta ^{k})\text{,\ \ \ }k=1,2,...,N.  \label{z}
\end{equation}%
The eigenvalue--eigenfunction problem is%
\begin{equation*}
\left| 
\begin{array}{ccc}
-\mu & 0 & 1 \\ 
1 & (u^{i}-\mu )\delta _{ik} & 0 \\ 
0 & \eta ^{i}\delta _{ik} & (u^{i}-\mu )\delta _{ik}%
\end{array}%
\right| \left| 
\begin{array}{c}
r \\ 
q^{i} \\ 
s^{i}%
\end{array}%
\right| =0.
\end{equation*}%
Thus,%
\begin{equation*}
q^{i}=\frac{r}{\mu -u^{i}}\text{, \ \ \ \ \ \ \ \ }s^{i}=\frac{\eta ^{i}}{%
(\mu -u^{i})^{2}}r\text{,}
\end{equation*}%
where $2N+1$ eigenvalues are solutions of the discriminant equation%
\begin{equation*}
\mu =\sum \frac{\eta ^{k}}{(\mu -u^{k})^{2}}.
\end{equation*}%
In accordance with the Tsarev observations, let us integrate this expression
once. Indeed,%
\begin{equation*}
\lambda =\frac{\mu ^{2}}{2}+A^{0}+\sum \frac{\eta ^{k}}{\mu -u^{k}}
\end{equation*}%
is the equation of the Riemann surface connected with the hydrodynamic type
system (\textbf{\ref{z}}). The factorized form of this equation%
\begin{equation}
\lambda =\frac{1}{2}\left( \mu -\sum u^{k}+\sum a^{k}\right) \frac{\overset{%
N+1}{\underset{n=1}{\prod }}(\mu -a^{n})}{\overset{N}{\underset{k=1}{\prod }}%
(\mu -u^{k})}  \label{tota}
\end{equation}%
yields $2N+1$ independent series of conservation laws%
\begin{eqnarray*}
\mu ^{(k)} &=&u^{k}+\lambda \eta ^{k}(\mathbf{u},\mathbf{a})+\lambda w^{k}(%
\mathbf{u},\mathbf{a})+...\text{, \ \ \ \ }k=1,2,...,N, \\
&& \\
\mu ^{(n)} &=&a^{n}+\lambda b^{n}(\mathbf{u},\mathbf{a})+\lambda c^{n}(%
\mathbf{u},\mathbf{a})+...\text{, \ \ \ \ }n=1,2,...,N+1,
\end{eqnarray*}%
whose coefficients can be found with the aid of the B\"{u}rmann--Lagrange
expansion (see (\textbf{\ref{ryad}})).

\textbf{Remark}: While the conservation law densities $\eta ^{k}$ are
coefficients of the first $N$ series, the function $A^{0}$ is a coefficient
of the Kruskal series for (\textbf{\ref{tota}}) at infinity $\lambda
\rightarrow \infty $, $\mu \rightarrow \infty $%
\begin{equation*}
\lambda =\frac{\mu ^{2}}{2}+A^{0}+\frac{A^{1}}{\mu }+...,
\end{equation*}%
where $A^{1}\equiv \Sigma \eta ^{n}$ (see (\textbf{\ref{third}}) and (%
\textbf{\ref{z}})).

\subsection{Hamiltonian formalism}

The Hamiltonian formalism of hydrodynamic type systems was established in 
\textbf{\cite{Dubr+Nov}} for a local case, and developed later for a
nonlocal case in \textbf{\cite{Fer+trans}}, \textbf{\cite{Fer+Mokh}}. The
generating function of conservation laws for the symmetric hydrodynamic type
systems (\textbf{\ref{1}}) is given by (\textbf{\ref{2}}). Suppose for
simplicity without lost of generality that the field variables $u^{k}$ are
flat coordinates. Then (\textbf{\ref{1}}) can be written in the \textit{%
symmetric} Hamiltonian form%
\begin{equation*}
u_{t}^{i}=\partial _{x}\left( \varepsilon _{i}\frac{\partial \mathbf{h}}{%
\partial u^{i}}+\gamma _{i}\underset{k\neq i}{\sum }\gamma _{k}\frac{%
\partial \mathbf{h}}{\partial u^{k}}\right) ,
\end{equation*}%
where the Hamiltonian density $\mathbf{h}$ is a symmetric expression under
an index permutation ($\varepsilon _{i}$ and $\gamma _{i}$ are some
constants). Thus, the generating function of commuting flows is given by%
\begin{equation*}
u_{\tau }^{i}=\partial _{x}\left( \varepsilon _{i}\frac{\partial p}{\partial
u^{i}}+\gamma _{i}\underset{k\neq i}{\sum }\gamma _{k}\frac{\partial p}{%
\partial u^{k}}\right) .
\end{equation*}%
Then infinitely many particular solutions are given by the generalized
hodograph method (\textbf{\ref{tot}})%
\begin{equation*}
x\delta _{k}^{i}+t\frac{\partial }{\partial u^{k}}\left( \varepsilon _{i}%
\frac{\partial \mathbf{h}}{\partial u^{i}}+\gamma _{i}\underset{m\neq i}{%
\sum }\gamma _{m}\frac{\partial \mathbf{h}}{\partial u^{m}}\right) =\overset{%
\infty }{\underset{n=1}{\sum }}\overset{N}{\underset{s=1}{\sum }}\sigma _{sn}%
\frac{\partial }{\partial u^{k}}\left( \varepsilon _{i}\frac{\partial 
\mathbf{h}_{n}^{(s)}}{\partial u^{i}}+\gamma _{i}\underset{m\neq i}{\sum }%
\gamma _{m}\frac{\partial \mathbf{h}_{n}^{(s)}}{\partial u^{m}}\right) ,
\end{equation*}%
where $\sigma _{sn}$ are arbitrary constants.

\textbf{Example}: \textit{The Hamiltonian exponential chromatography} (%
\textbf{\ref{exp}})%
\begin{equation*}
u_{t}^{i}=\partial _{x}\frac{e^{u^{i}}}{1+\sum \gamma _{k}e^{u^{k}}}\text{,
\ \ \ \ \ }i=1,2,...,N
\end{equation*}%
is connected with the equation of the Riemann surface (\textbf{\ref{rexp}})
by the Gibbons equation%
\begin{equation*}
\lambda _{t}-\frac{e^{p}}{\Delta }\lambda _{x}=\frac{\partial \lambda }{%
\partial p}\left[ p_{t}-\partial _{x}\frac{e^{p}}{\Delta }\right] .
\end{equation*}%
The Kruskal series of conservation law densities can be found by the
application of the B\"{u}rmann--Lagrange series (\textbf{\ref{ryad}}) at the
vicinity $\lambda \rightarrow 0,q\rightarrow 0$, where $q=\exp (-p)$. In
this case the coefficients $q_{n}$ of the inverse series%
\begin{equation*}
q=q_{1}\lambda +q_{2}\lambda ^{2}+q_{3}\lambda ^{3}+q_{4}\lambda ^{4}+...
\end{equation*}%
are determined by%
\begin{equation*}
q_{n}=\frac{1}{n!}\underset{q\rightarrow 0}{\lim }\frac{d^{n-1}}{dq^{n-1}}%
\left( \frac{q}{q-\sum \gamma _{k}\ln (1-qe^{u^{k}})}\right) ^{n}\text{, \ \
\ \ \ }n=1,2,...
\end{equation*}%
Then the Kruskal conservation law densities can be found from $p=-\ln q$%
\begin{equation*}
\tilde{p}=p+\ln \lambda =-\ln [1+q_{2}\lambda +q_{3}\lambda
^{2}+q_{4}\lambda ^{3}+...].
\end{equation*}%
For instance, $p_{1}=\ln \Delta $, $p_{2}=\Delta ^{-4}\Sigma \gamma
_{n}e^{2u^{n}}$.

The Kruskal series can be extended in the \textit{opposite} direction $%
\lambda \rightarrow \infty ,q\rightarrow \infty $. At first, the equation of
the Riemann surface (\textbf{\ref{rexp}}) must be replaced by (cf. (\textbf{%
\ref{wat}}))%
\begin{equation*}
\lambda -\sum \gamma _{k}\ln \lambda =s-\sum \gamma _{k}\ln (s+\sum \gamma
_{m}u^{m}-e^{-u^{k}}),
\end{equation*}%
where $s=\exp (-p)-\Sigma \gamma _{m}u^{m}$. Then the Kruskal conservation
law densities $p_{-k}$ can be found exactly as for the Benney hydrodynamic
chain (\textbf{\ref{bm}}) (see (\textbf{\ref{wat}})) by substitution (cf. (%
\textbf{\ref{chu}}))%
\begin{equation*}
s=\lambda +\frac{C_{1}}{\lambda }+\frac{C_{2}}{\lambda ^{2}}+\frac{C_{3}}{%
\lambda ^{3}}+...
\end{equation*}%
in the above formula. Then $p_{-k}$ are coefficients of the series%
\begin{equation*}
\tilde{p}=p+\ln \lambda =\ln \left( 1+\frac{1}{\lambda }\sum \gamma
_{m}u^{m}+\frac{C_{1}}{\lambda ^{2}}+\frac{C_{2}}{\lambda ^{3}}+\frac{C_{3}}{%
\lambda ^{4}}+...\right) .
\end{equation*}%
For instance, $p_{-1}=\Sigma \gamma _{m}u^{m}$, $p_{-2}=\Sigma \gamma
_{m}e^{-u^{m}}+(\Sigma \gamma _{m}u^{m})^{2}/2$.

Any semi-Hamiltonian hydrodynamic type system has $N$ series of conservation
laws. The equation of the Riemann surface (\textbf{\ref{rexp}}) can be
written in the form%
\begin{equation*}
\lambda =(e^{-p}-e^{-u^{i}})\exp \left( u^{i}-\frac{e^{-p}}{\gamma _{i}}+%
\underset{k\neq i}{\sum }\frac{\gamma _{k}}{\gamma _{i}}\ln
(e^{u^{k}-p}-1)\right) \text{, \ \ \ \ \ }p\rightarrow u^{i}.
\end{equation*}%
Then the coefficients $q_{k}$ of the B\"{u}rmann--Lagrange series are given
by (\textbf{\ref{ryad}})%
\begin{equation*}
q_{n}=\frac{1}{n!}\underset{q\rightarrow \exp (-u^{i})}{\lim }\frac{d^{n-1}}{%
dq^{n-1}}\left( \frac{q-e^{-u^{i}}}{\lambda (q)}\right) ^{n}\text{, \ \ \ \
\ }n=1,2,...,
\end{equation*}%
where $q=\exp (-p)$. Finally, $N$ series of conservation law densities $%
p^{(i)}$ can be found from the series%
\begin{equation*}
\tilde{p}^{(i)}=p^{(i)}+\ln \lambda =-\ln [q_{1}+q_{2}\lambda +q_{3}\lambda
^{2}+q_{4}\lambda ^{3}+...].
\end{equation*}%
For instance, the first $N$ nontrivial conservation law densities are%
\begin{equation}
p_{1}^{(i)}=e^{-u^{i}}-\underset{k\neq i}{\sum }\gamma _{k}\ln
(e^{u^{k}-u^{i}}-1).  \label{odin}
\end{equation}

Commuting flows of this Hamiltonian chromatography system are%
\begin{eqnarray*}
u_{t^{k}}^{i} &=&\frac{1}{\gamma _{i}}\partial _{x}\frac{\partial \mathbf{h}%
_{k}}{\partial u^{i}}\text{, \ \ \ \ \ \ }k=0,\pm 1,\pm 2,...\text{,} \\
&& \\
u_{t^{k,n}}^{i} &=&\frac{1}{\gamma _{i}}\partial _{x}\frac{\partial \mathbf{h%
}_{k,n}}{\partial u^{i}}\text{, \ \ \ \ }k=1,2,...\text{, \ \ \ \ }%
n=1,2,...,N\text{,}
\end{eqnarray*}%
where $\mathbf{h}_{k}$ are Kruskal conservation law densities and $\mathbf{h}%
_{k,n}$ are $N$ series of conservation law densities. The first $N$
commuting flows%
\begin{equation*}
u_{t^{1,k}}^{k}=\frac{1}{\gamma _{k}}\partial _{x}\left( \underset{n\neq i}{%
\sum }\gamma _{n}\frac{e^{u^{n}}}{e^{u^{n}}-e^{u^{k}}}-e^{-u^{k}}\right) 
\text{, \ \ \ \ \ \ \ }u_{t^{1,k}}^{i}=\partial _{x}\frac{e^{u^{i}}}{%
e^{u^{k}}-e^{u^{i}}}\text{, \ \ \ }i\neq k
\end{equation*}%
are given by (\textbf{\ref{odin}}).

\subsection{\textit{Mirrored} curvilinear conjugate coordinate systems}

If the semi-Hamiltonian property (\textbf{\ref{semi}}) is valid, then the
diagonal metric $g_{ii}$ can be introduced by%
\begin{equation}
\frac{\partial _{k}\mu ^{i}}{\mu ^{k}-\mu ^{i}}=\partial _{k}\ln H_{i}\text{%
, \ \ \ \ }i\neq k.  \label{one}
\end{equation}%
Following G. Darboux (see \textbf{\cite{Darboux}}), let us introduce the
rotation coefficients of the conjugate curvilinear coordinate nets%
\begin{equation}
\beta _{ik}=\frac{\partial _{i}H_{k}}{H_{i}}\text{, \ \ \ \ }i\neq k.
\label{two}
\end{equation}%
Then any solution $\tilde{H}_{k}$ of the linear problem (see \textbf{\cite%
{Tsar}})%
\begin{equation}
\partial _{i}\tilde{H}_{k}=\beta _{ik}\tilde{H}_{i}\text{, \ \ \ \ }i\neq k
\label{lin}
\end{equation}%
determines commuting flow to (\textbf{\ref{ri}}), where characteristic
velocities $w^{i}$ are connected with $\tilde{H}_{i}$ by the Combescure
transformation $w^{i}=\tilde{H}_{i}/H_{i}$. Any solution of the conjugate
linear problem%
\begin{equation}
\partial _{i}\psi _{k}=\beta _{ki}\psi _{i}\text{, \ \ \ \ }i\neq k
\label{gug}
\end{equation}%
determines conservation law density%
\begin{equation}
\partial _{i}a=\psi _{i}H_{i}\text{.}  \label{cha}
\end{equation}

\textbf{Definition 2} \textbf{\cite{Maks+Tsar}}: \textit{The conjugate
curvilinear coordinate net determined by symmetric rotation coefficients} $%
\beta _{ik}=\beta _{ki}$ \textit{is called the Egorov conjugate curvilinear
coordinate net}.

\textbf{Theorem 2} \textbf{\cite{Maks+Tsar}}: \textit{If the integrable
hydrodynamic type system} (\textbf{\ref{ri}}) \textit{has the couple of
conservation laws}%
\begin{equation}
a_{t}=b_{x}\text{, \ \ \ \ \ \ \ }b_{t}=c_{x},  \label{egor}
\end{equation}%
\textit{then corresponding conjugate curvilinear coordinate net is Egorov}.

\textbf{Remark \cite{Maks+Tsar}}: This couple is unique for each given Lame
coefficients $H_{k}$.

\textbf{Corollary \cite{Maks+Tsar}}: Any commuting flow (\textbf{\ref{com}})
has similar couple of conservation laws (\textbf{\ref{egor}})%
\begin{equation*}
a_{y}=h_{x}\text{, \ \ \ \ \ }h_{y}=f_{x},
\end{equation*}%
where $a$ is the unique potential of the Egorov metric for all commuting
flows.

\textbf{Definition 3}: \textit{Two conjugate curvilinear coordinate nets
determined by the rotation coefficients} $\beta _{ik}$ \textit{and} $\bar{%
\beta}_{ik}=\beta _{ki}$ \textit{is called \textbf{mirrored} conjugate
curvilinear coordinate nets}.

\textbf{Theorem 3}: \textit{If one integrable hydrodynamic type system} (%
\textbf{\ref{ri}}) \textit{has the conservation law}%
\begin{equation}
a_{t}=b_{x}  \label{first}
\end{equation}%
\textit{such that another integrable hydrodynamic type system}%
\begin{equation}
r_{y}^{i}=\tilde{\mu}^{i}(\mathbf{r})r_{z}^{i}  \label{tri}
\end{equation}%
\textit{has the couple of conservation laws}%
\begin{equation}
a_{y}=c_{z}\text{, \ \ \ \ \ \ \ }b_{y}=B_{z},  \label{sec}
\end{equation}%
\textit{then corresponding conjugate curvilinear coordinate nets are mirrored%
}.

\textbf{Proof}: The characteristic velocities $\upsilon ^{i}$ of the
hydrodynamic type system (\textbf{\ref{ri}}) are (see (\textbf{\ref{cha}}))%
\begin{equation*}
\mu ^{i}=\frac{H_{i}^{(1)}}{H_{i}}=\frac{\psi _{i}H_{i}^{(1)}}{\psi _{i}H_{i}%
}=\frac{\partial _{i}b}{\partial _{i}a}\text{,}
\end{equation*}%
where $H_{i}^{(1)}$ is a solution of the linear system (\textbf{\ref{lin}}).
Let us consider such hydrodynamic type system (\textbf{\ref{tri}}), whose
characteristic velocities $\tilde{\mu}^{i}$ are (see (\textbf{\ref{cha}}))%
\begin{equation*}
\tilde{\mu}^{i}=\frac{\psi _{i}^{(1)}}{\psi _{i}}=\frac{H_{i}\psi _{i}^{(1)}%
}{H_{i}\psi _{i}}=\frac{\partial _{i}c}{\partial _{i}a}=\frac{%
H_{i}^{(1)}\psi _{i}^{(1)}}{H_{i}^{(1)}\psi _{i}}=\frac{\partial _{i}B}{%
\partial _{i}b}\text{,}
\end{equation*}%
where $\psi _{i}^{(1)}$ is a solution of the conjugate linear system (%
\textbf{\ref{gug}}). The rotation coefficients can be found in two steps (%
\textbf{\ref{one}}) and (\textbf{\ref{two}}). Indeed, $\bar{\beta}%
_{ik}=\beta _{ki}$.

\textbf{Remark}: The construction described above is symmetric. Thus, the
second conservation law of the first hydrodynamic type system (\textbf{\ref%
{ri}})%
\begin{equation*}
c_{t}=C_{x}
\end{equation*}%
is given by quadratures%
\begin{equation*}
dc=\sum \psi _{i}^{(1)}H_{i}dr^{i}\text{, \ \ \ \ \ }dC=\sum \psi
_{i}^{(1)}H_{i}^{(1)}dr^{i}.
\end{equation*}

\textbf{Corollary}: Any conservation law%
\begin{equation*}
P_{y}=Q_{z}
\end{equation*}%
of the hydrodynamic type system (\textbf{\ref{tri}}) determines the
corresponding commuting flow in the conservative form%
\begin{equation}
a_{\tau }=P_{x}  \label{zakon}
\end{equation}%
or in the Riemann invariants (see(\textbf{\ref{com}}))%
\begin{equation}
r_{\tau }^{i}=\frac{H_{i}^{(2)}}{H_{i}}r_{x}^{i}  \label{rimm}
\end{equation}%
of the hydrodynamic type system (\textbf{\ref{ri}}), where $H_{i}^{(2)}$ is
some solution of the linear system (\textbf{\ref{lin}}), $\partial
_{i}P=H_{i}^{(2)}\psi _{i}$ and $\partial _{i}Q=H_{i}^{(2)}\psi _{i}^{(1)}$.

Thus, if two symmetric hydrodynamic type systems are related by the above
link, then the generating function of the second hydrodynamic type system (%
\textbf{\ref{tri}}) determines the generating function of commuting flows
for the first hydrodynamic type system (\textbf{\ref{ri}}).

\subsection{Chromatography system}

The integrable chromatography system (\textbf{\ref{ful}}) has the couple of
conservation laws%
\begin{equation*}
\partial _{t}\left[ \sum \gamma _{k}(u^{k})^{\beta -\beta \varepsilon +1}%
\right] =(\beta \varepsilon -\beta -1)\partial _{x}\Delta ^{-\varepsilon },
\end{equation*}%
\begin{equation*}
\partial _{t}\Delta ^{1/\beta }=\partial _{x}\left[ \frac{\beta \varepsilon 
}{\beta \varepsilon +\beta -1}\Delta ^{\frac{1-\beta -\beta \varepsilon }{%
\beta }}\sum \gamma _{n}(u^{n})^{\beta \varepsilon +\beta -1}\right] .
\end{equation*}

\textbf{1}. The \textit{Egorov} sub-case. If $\beta \varepsilon =-1$, then
this is the Egorov hydrodynamic type system (see (\textbf{\ref{egor}}))%
\begin{equation*}
u_{t}^{i}=\partial _{x}\frac{[1+\sum \gamma _{k}(u^{k})^{\beta }]^{1/\beta }%
}{u^{i}}\text{, \ \ \ \ \ }i=1,2,...,N
\end{equation*}%
where the potential of the Egorov metric (see \textbf{\cite{Maks+Tsar}}) is%
\begin{equation*}
a=\sum \gamma _{k}(u^{k})^{\beta +2}.
\end{equation*}

\textbf{2}. The \textit{local Hamiltonian} sub-case. If $\beta
(1-\varepsilon )=1$, then chromatography system (\textbf{\ref{ful}}) can be
written in the Hamiltonian form (\textbf{\ref{ham}}). If $\beta \rightarrow
\infty $, the exponential chromatography (\textbf{\ref{exp}}) ($\varepsilon
=1$) also has the Hamiltonian form (\textbf{\ref{ham}}), where the
Hamiltonian density is $\mathbf{h}=\ln \Delta $.

\textbf{3}. The \textit{nonlocal Hamiltonian} sub-case (associated with
constant curvature metric, see \textbf{\cite{Fer+Mokh}}, \textbf{\cite%
{Maks+cc}}). If $\beta =2$, then the conservation law density%
\begin{equation*}
q=1-\sqrt{1+\sum \gamma _{k}(u^{k})^{2}}
\end{equation*}%
is the momentum density (see \textbf{\cite{Maks+cc}}) of the nonlocal
Hamiltonian structure%
\begin{equation*}
u_{t}^{i}=\partial _{x}\left( (\bar{g}^{ik}-u^{i}u^{k})\frac{\partial 
\mathbf{\tilde{h}}}{\partial u^{k}}+u^{i}\mathbf{\tilde{h}}\right) ,
\end{equation*}%
where diagonal matrix elements $\bar{g}_{ik}=-\gamma _{i}\delta _{ik}$
(where $\delta _{ik}$ is the Kronecker symbol) and the Hamiltonian density%
\begin{equation*}
\mathbf{\tilde{h}}=-\frac{\Delta ^{-\varepsilon }}{2\varepsilon +1}\sum
\gamma _{k}(u^{k})^{2\varepsilon +1}\text{.}
\end{equation*}

\textbf{4}. The general (\textit{mirrored}) sub-case. If two hydrodynamic
type systems (\textbf{\ref{ful}}) are related via mirrored conjugate
curvilinear coordinate nets, then they must have one common conservation law
density (see (\textbf{\ref{first}}) and (\textbf{\ref{sec}})). Let us prove
that such conservation law density is%
\begin{equation}
a=\sum \gamma _{k}(u^{k})^{\beta -\beta \varepsilon +1}.  \label{den}
\end{equation}%
If the another chromatography system (\textbf{\ref{ful}})%
\begin{equation}
\tilde{u}_{y}^{i}=\partial _{z}\frac{(\tilde{u}^{i})^{\tilde{\beta}\tilde{%
\varepsilon}}}{[1+\sum \gamma _{k}(\tilde{u}^{k})^{\tilde{\beta}}]^{\tilde{%
\varepsilon}}}\text{, \ \ \ \ \ }i=1,2,...,N  \label{drug}
\end{equation}%
has the same conservation law density%
\begin{equation*}
a=\sum \gamma _{k}(\tilde{u}^{k})^{\tilde{\beta}-\tilde{\beta}\tilde{%
\varepsilon}+1},
\end{equation*}%
then transformation $u\rightarrow \tilde{u}$, $\beta \rightarrow \tilde{\beta%
}$, $\varepsilon \rightarrow \tilde{\varepsilon}$ can be found from two
other obvious assumptions (see (\textbf{\ref{first}}) and (\textbf{\ref{sec}}%
))%
\begin{equation*}
u^{i}=(\tilde{u}^{i})^{\delta }\text{, \ \ \ \ \ }\Delta ^{-\varepsilon }=%
\tilde{\Delta}^{1/\tilde{\beta}}.
\end{equation*}

\textbf{Result}:%
\begin{equation*}
\delta =-\frac{1}{\beta \varepsilon }\text{, \ \ \ \ }\tilde{\varepsilon}=-%
\frac{1}{\beta }\text{, \ \ \ \ }\tilde{\beta}=-\frac{1}{\varepsilon }.
\end{equation*}

Since the second chromatography system (\textbf{\ref{drug}}) has the
generating function of conservation laws%
\begin{equation*}
P_{y}=\partial _{z}\frac{P^{\frac{1}{\beta \varepsilon }}}{[1+\sum \gamma
_{k}(u^{k})^{\beta }]^{-1/\beta }}\text{,}
\end{equation*}%
then the generating function of commuting flows is given by (\textbf{\ref%
{zakon}})%
\begin{equation}
a_{\tau }=\partial _{x}p^{-\beta \varepsilon },  \label{full}
\end{equation}%
where the relationship%
\begin{equation*}
P=p^{-\beta \varepsilon }
\end{equation*}%
is obtained by a comparison the equations of the Riemann surface (which are
equivalent) for both chromatography systems.

The integrable chromatography system (\textbf{\ref{ful}}) in the Riemann
invariants has the form (\textbf{\ref{rim}})%
\begin{equation*}
r_{t}^{i}=\beta \varepsilon \frac{(p^{i})^{\beta \varepsilon -1}}{\Delta
^{\varepsilon }}r_{x}^{i}
\end{equation*}%
Since this hydrodynamic type system has two conservation laws%
\begin{equation*}
p_{t}=\partial _{x}\frac{p^{\beta \varepsilon }}{\Delta ^{\varepsilon }}%
\text{, \ \ \ \ \ \ }a_{t}=(\beta \varepsilon -\beta -1)\partial _{x}\Delta
^{-\varepsilon },
\end{equation*}%
then one can obtain, respectively%
\begin{equation*}
\partial _{i}p=\frac{p^{\beta \varepsilon }}{p^{\beta \varepsilon
-1}-(p^{i})^{\beta \varepsilon -1}}\frac{\partial _{i}\ln \Delta }{\beta }%
\text{, \ \ \ \ \ }\partial _{i}a=\frac{1+\beta -\beta \varepsilon }{\beta }%
(p^{i})^{1-\beta \varepsilon }\partial _{i}\ln \Delta .
\end{equation*}%
Thus, the generating function of commuting flows (\textbf{\ref{full}}) in
the Riemann invariants (\textbf{\ref{rimm}})%
\begin{equation*}
r_{\tau }^{i}=\sigma \frac{(p^{i})^{\beta \varepsilon -1}}{p[(p^{i})^{\beta
\varepsilon -1}-p^{\beta \varepsilon -1}]}r_{x}^{i}
\end{equation*}%
is connected with the Gibbons equation%
\begin{eqnarray*}
&&\lambda _{\tau (\zeta )}-\frac{\beta \varepsilon }{1+\beta -\beta
\varepsilon }\frac{(p^{i})^{\beta \varepsilon -1}}{p(\zeta )[(p^{i})^{\beta
\varepsilon -1}-p^{\beta \varepsilon -1}(\zeta )]}\lambda _{x} \\
&& \\
&=&\frac{\partial \lambda }{\partial p(\lambda )}\left[ \partial _{\tau
(\zeta )}p(\lambda )+\frac{\beta \varepsilon }{(\beta \varepsilon
-1)(1+\beta -\beta \varepsilon )}\partial _{x}\left[ \left( \frac{p(\lambda )%
}{p(\zeta )}\right) ^{\beta \varepsilon }F\left( 1,\sigma ,\sigma +1,\frac{%
p^{\beta \varepsilon -1}(\lambda )}{p^{\beta \varepsilon -1}(\zeta )}\right) %
\right] \right] ,
\end{eqnarray*}%
where we use the notation%
\begin{equation*}
\sigma =\frac{\beta \varepsilon }{\beta \varepsilon -1}.
\end{equation*}%
Then the generating function of commuting flows written via the physical
field variables $u^{k}$ (cf. (\textbf{\ref{ful}})) is%
\begin{equation*}
u_{\tau (\zeta )}^{i}=-\frac{\beta \varepsilon }{(\beta \varepsilon
-1)(1+\beta -\beta \varepsilon )}\partial _{x}\left[ \left( \frac{u^{i}}{%
p(\zeta )}\right) ^{\beta \varepsilon }F\left( 1,\sigma ,\sigma +1,\frac{%
(u^{i})^{\beta \varepsilon -1}}{p^{\beta \varepsilon -1}(\zeta )}\right) %
\right] .
\end{equation*}%
Substituting the formal series $\partial _{\tau (\zeta )}=\partial
_{t_{0}^{i}}+\zeta \partial _{t_{1}^{i}}+\zeta ^{2}\partial _{t_{2}^{i}}+...$
and respectively the generating function of conservation law densities
expanded in the series (\textbf{\ref{ser}}), one can obtain infinitely many
generating functions of conservation laws for all commuting flows. For
instance, the first $N$ such functions are%
\begin{equation*}
p_{t_{0}^{i}}=\partial _{x}\left[ \left( \frac{p}{u^{i}}\right) ^{\beta
\varepsilon }F\left( 1,\sigma ,\sigma +1,\frac{p^{\beta \varepsilon -1}}{%
(u^{i})^{\beta \varepsilon -1}}\right) \right] .
\end{equation*}

\textbf{Remark}: Suppose that some $N$ component hydrodynamic type system
contains $N-1$ equations (see the above generation function of conservation
laws)%
\begin{equation*}
u_{t_{0}^{i}}^{k}=\partial _{x}\left[ \left( \frac{u^{k}}{u^{i}}\right)
^{\beta \varepsilon }F\left( 1,\sigma ,\sigma +1,\frac{(u^{k})^{\beta
\varepsilon -1}}{(u^{i})^{\beta \varepsilon -1}}\right) \right] \text{, \ \
\ \ \ }k\neq i.
\end{equation*}%
Then such hydrodynamic type system is integrable if its $N$th equation
satisfies some extra conditions. This hydrodynamic type system is not
symmetric like (\textbf{\ref{1}}). However, the algebro-geometric approach
still is valid.

Thus, if any given hydrodynamic type system contains $N-1$ equations%
\begin{equation*}
u_{t}^{k}=\partial _{x}\psi \left( \mathbf{u};\frac{u^{k}}{u^{1}}\right) 
\text{, \ \ \ \ }k\neq 1,
\end{equation*}%
then one should substitute the Taylor series (\textbf{\ref{ser}}) in the
generating function of conservation laws%
\begin{equation*}
p_{t}=\partial _{x}\psi \left( \mathbf{u};\frac{p}{u^{1}}\right) \text{.}
\end{equation*}%
The $N$th equation will be obtained by the limit (see such example at the
end of the sub-section ``Hamiltonian formalism'')%
\begin{equation*}
u_{t}^{1}=\partial _{x}\left[ \underset{\varepsilon \rightarrow 0}{\lim }%
\psi \left( \mathbf{u};1+\varepsilon \frac{h_{1}^{1}(\mathbf{u})}{u^{1}}%
\right) \right] .
\end{equation*}%
All other computations are exactly as in the symmetric case.

\subsection{Reciprocal transformations}

The concept ``reciprocal transformation'' was introduced by S.A. Chaplygin
(see, for instance, \textbf{\cite{Rogers}} and \textbf{\cite{Yanenko}})%
\begin{equation*}
dz=A(\mathbf{u})dx+B(\mathbf{u})dt\text{, \ \ \ \ \ \ \ }dy=C(\mathbf{u}%
)dx+D(\mathbf{u})dt,
\end{equation*}%
where two arbitrary conservation laws $A_{t}=B_{x}$ and $C_{t}=D_{x}$
preserve the gas dynamic equations changing adiabatic index only. The gas
dynamics is the first known example in the theory of integrable hydrodynamic
type systems. In this sub-section we show that the integrable chromatography
system (\textbf{\ref{ful}}) is invariant under couple of different
reciprocal transformations. Thus, infinite sets of such systems are related
by a \textit{chain} of reciprocal transformations, which are described below.

\textbf{1}. The first such reciprocal transformation is very simple%
\begin{equation*}
dz=dt\text{, \ \ \ \ \ \ \ }dy=dx.
\end{equation*}%
Then the integrable chromatography system (\textbf{\ref{ful}}) reduces to
the chromatography system%
\begin{equation*}
w_{y}^{i}=\partial _{z}\frac{(w^{i})^{\frac{1}{\beta \varepsilon }}}{[1-\sum
\gamma _{k}(w^{k})^{1/\varepsilon }]^{1/\beta }}\text{, \ \ \ \ \ }%
i=1,2,...,N,
\end{equation*}%
where%
\begin{equation*}
u^{i}=(w^{i})^{\frac{1}{\beta \varepsilon }}\bar{\Delta}^{-1/\beta }\text{,
\ \ \ \ \ \ \ \ }\bar{\Delta}=\Delta ^{-1}.
\end{equation*}%
Thus, two chromatography systems (\textbf{\ref{ful}}) ($\beta ,\varepsilon
,\gamma _{k}$) and ($1/\varepsilon ,1/\beta ,-\gamma _{k}$) are related by
the transformation $x\leftrightarrow t$.

\textbf{Remark}: The first conservation law%
\begin{equation*}
\partial _{t}\left[ \sum \gamma _{k}(u^{k})^{\beta -\beta \varepsilon +1}%
\right] =(\beta \varepsilon -\beta -1)\partial _{x}\Delta ^{-\varepsilon }
\end{equation*}%
transforms into the second conservation law%
\begin{equation*}
\partial _{y}\bar{\Delta}^{1/\bar{\beta}}=\partial _{z}\left[ \frac{\bar{%
\beta}\bar{\varepsilon}}{\bar{\beta}\bar{\varepsilon}+\bar{\beta}-1}\bar{%
\Delta}^{\frac{1-\bar{\beta}-\bar{\beta}\bar{\varepsilon}}{\bar{\beta}}}\sum 
\bar{\gamma}_{n}(w^{n})^{\bar{\beta}\bar{\varepsilon}+\bar{\beta}-1}\right]
\end{equation*}%
and vice versa.

\textbf{2}. The reciprocal transformation%
\begin{equation*}
dz=\Delta ^{1/\beta }dx+\left[ \frac{\beta \varepsilon }{\beta \varepsilon
+\beta -1}\Delta ^{\frac{1-\beta -\beta \varepsilon }{\beta }}\sum \gamma
_{n}(u^{n})^{\beta \varepsilon +\beta -1}\right] dt\text{, \ \ \ \ \ }%
dy=\beta \varepsilon dt
\end{equation*}%
connects the chromatography system (\textbf{\ref{ful}}) and another
symmetric system%
\begin{equation}
\upsilon _{y}^{i}=\partial _{z}\left( \frac{(\upsilon ^{i})^{\beta
\varepsilon }}{\beta \varepsilon }-\frac{\upsilon ^{i}}{\beta \varepsilon
+\beta -1}\sum \gamma _{n}(\upsilon ^{n})^{\beta \varepsilon +\beta
-1}\right) \text{, \ \ \ \ \ }i=1,2,...,N,  \label{k}
\end{equation}%
where $\upsilon ^{i}=u^{i}\Delta ^{-1/\beta }$.

However, the chromatography system (\textbf{\ref{ful}}) has the commuting
flow%
\begin{equation}
u_{t^{1}}^{i}=\partial _{x}\left[ \frac{(u^{i})^{2-\beta \varepsilon }}{%
2-\beta \varepsilon }+\frac{u^{i}}{\beta -\beta \varepsilon +1}\sum \gamma
_{n}(u^{n})^{\beta -\beta \varepsilon +1}\right] .  \label{poli}
\end{equation}%
It is easy to verify by the compatibility conditions $\partial
_{t^{1}}(u_{t}^{i})=\partial _{t}(u_{t^{1}}^{i})$. Then the hydrodynamic
type system (\textbf{\ref{k}}) has the commuting flow%
\begin{equation*}
\upsilon _{t}^{i}=\partial _{x}\frac{(\upsilon ^{i})^{2-\beta \varepsilon }}{%
[1-\sum \gamma _{k}(\upsilon ^{k})^{\beta }]^{2/\beta -\varepsilon }}\text{,
\ \ \ \ \ }i=1,2,...,N
\end{equation*}%
which is the chromatography system (\textbf{\ref{ful}}) again. Thus, two
chromatography systems (\textbf{\ref{ful}}) ($\beta ,\varepsilon ,\gamma
_{k} $) and ($\beta ,2/\beta -\varepsilon ,-\gamma _{k}$) are related by the
above reciprocal transformation, where $\bar{\Delta}=\Delta ^{-1}=1-\Sigma
\gamma _{n}(\upsilon ^{n})^{\beta }$.

Applying both reciprocal transformations iteratively, one can construct a
link between the chromatography systems (\textbf{\ref{ful}}) with the
distinct indexes $\beta $ and $\varepsilon $.

\section{\textit{Homogeneous} hydrodynamic type systems}

Another hydrodynamic type system%
\begin{equation*}
u_{t}^{i}=\partial _{x}\left( (u^{i})^{\beta }\prod (u^{n})^{\gamma
_{n}}\right)
\end{equation*}%
arising in the chromatography (see formula \textbf{6} in \textbf{\cite%
{Fer+Tsar}}, $\beta $ and $\gamma _{n}$ are arbitrary constants) is
invariant under scaling of field variables $u^{k}\rightarrow cu^{k}$ (and
appropriate scaling of independent variable $t$ or $x$). We call such
hydrodynamic type systems as \textit{homogeneous}. Plenty physically
interested hydrodynamic type systems belong to this class.

The existence of the corresponding generating function of conservation laws%
\begin{equation*}
p_{t}=\partial _{x}\left( p^{\beta }\prod (u^{n})^{\gamma _{n}}\right)
\end{equation*}%
yields (see (\textbf{\ref{9}}))%
\begin{equation*}
\frac{\partial p}{\partial u^{i}}=\frac{\gamma _{i}p^{\beta }}{\beta
u^{i}[p^{\beta -1}-(u^{i})^{\beta -1}]}\left( \frac{1}{\beta }\sum \frac{%
\gamma _{k}(u^{k})^{\beta -1}}{p^{\beta -1}-(u^{i})^{\beta -1}}-1\right)
^{-1}.
\end{equation*}%
Also, the equation of the Riemann surface $\lambda (\mathbf{u},p)$ must be
invariant under \textit{extended} scaling $u^{k}\rightarrow cu^{k}$, $%
p\rightarrow cp$. Since (see (\textbf{\ref{7}}))%
\begin{equation*}
\frac{\partial \lambda }{\partial u^{i}}=\frac{\gamma _{i}p^{\beta }}{\beta
u^{i}[p^{\beta -1}-(u^{i})^{\beta -1}]}\left( 1-\frac{1}{\beta }\sum \frac{%
\gamma _{k}(u^{k})^{\beta -1}}{p^{\beta -1}-(u^{i})^{\beta -1}}\right) ^{-1}%
\frac{\partial \lambda }{\partial p},
\end{equation*}%
then the equation of the Riemann surface $\lambda (\mathbf{u},p)$ can be
found in quadratures%
\begin{equation*}
\ln \lambda =\ln p+\frac{1}{(\beta -1)(\beta +\sum \gamma _{m})}\sum \gamma
_{k}\ln \frac{(u^{k})^{\beta -1}}{p^{\beta -1}-(u^{k})^{\beta -1}},
\end{equation*}%
where we used the Euler theorem%
\begin{equation*}
\lambda =p\lambda _{p}+\sum u^{k}\lambda _{k}.
\end{equation*}%
Since the Gibbons equation (\textbf{\ref{3}}) is invariant under point
transformation $\lambda \rightarrow \tilde{\lambda}(\lambda )$, then the
above equation of the Riemann surface $\lambda (\mathbf{u},p)$ can be
written in the form%
\begin{equation*}
\lambda =q^{\beta +\sum \gamma _{m}}\prod \left( 1-\frac{q}{w^{k}}\right)
^{-\gamma _{k}},
\end{equation*}%
where $q=p^{\beta -1}$,\ $w^{k}=(u^{k})^{\beta -1}$.

\subsection{The Kodama hydrodynamic type system}

The Kodama hydrodynamic type system (see \textbf{\cite{Kod+water}})%
\begin{equation*}
a_{t}^{k}=\partial _{x}\left[ \frac{1}{2}\underset{m=1}{\overset{k}{\sum }}%
a^{m}a^{k+1-m}+\delta ^{k,1}a^{N}\right] \text{, \ \ \ \ \ \ }k=1,2,...,N,
\end{equation*}%
where $\delta ^{ik}$ is the Kronecker symbol, is a homogeneous (The Euler
operator is $\hat{E}=\Sigma (N+k-2)a^{k}\partial _{k}$) hydrodynamic
reduction of the Benney hydrodynamic chain (\textbf{\ref{bm}}). Thus, the
corresponding generating function of conservation laws is (\textbf{\ref{con}}%
) (the generating function of conservation laws for any integrable
hydrodynamic chain is unique for all its hydrodynamic reductions). The
Kodama hydrodynamic type system can be obtained from the above generating
function by substitution of the Taylor series%
\begin{equation*}
p=a^{1}+\lambda a^{2}+\lambda ^{2}a^{3}+...+\lambda ^{N-1}a^{N}+\lambda
^{N}h_{1}(\mathbf{a})+\lambda ^{N+1}h_{2}(\mathbf{a})+\lambda ^{N+2}h_{3}(%
\mathbf{a})+...
\end{equation*}%
Then $A^{0}\equiv a^{N}$ and $h_{k}(\mathbf{a})$ are some polynomial
conservation law densities. For instance, $\mathbf{h}_{1}(\mathbf{a})=\Sigma
a^{k}a^{N+1-k}/2$ is a momentum, $\mathbf{h}_{2}(\mathbf{a})$ is the
Hamiltonian, where the Kodama hydrodynamic type system has bi-Hamiltonian
structure, and the first of them is%
\begin{equation*}
a_{t}^{k}=\partial _{x}\frac{\partial \mathbf{h}_{2}}{\partial a^{N+1-k}}%
\text{, \ \ \ \ \ \ }k=1,2,...,N.
\end{equation*}%
Without lost of generality let us restrict our consideration on the three
component case%
\begin{equation*}
u_{t}=\partial _{x}\left( \frac{u^{2}}{2}+w\right) \text{, \ \ \ \ }\upsilon
_{t}=\partial _{x}(u\upsilon )\text{, \ \ \ \ }w_{t}=\partial _{x}\left( uw+%
\frac{\upsilon ^{2}}{2}\right) .
\end{equation*}%
Thus, the Gibbons equation (see (\textbf{\ref{3}}) and (\textbf{\ref{con}}))
is%
\begin{equation*}
\lambda _{t}-p\lambda _{x}=\frac{\partial \lambda }{\partial p}\left[
p_{t}-\partial _{x}\left( \frac{p^{2}}{2}+w\right) \right] .
\end{equation*}%
Then one has%
\begin{equation*}
\lambda _{u}=\frac{w(p-u)+\upsilon ^{2}}{\Delta }\lambda _{p}\text{, \ \ \ }%
\lambda _{\upsilon }=\frac{\upsilon (p-u)}{\Delta }\lambda _{p}\text{, \ \ \
\ }\lambda _{w}=\left[ \frac{(p-u)^{3}}{\Delta }-1\right] \lambda _{p}\text{,%
}
\end{equation*}%
where%
\begin{equation*}
\Delta =(p-u)^{3}-w(p-u)-\upsilon ^{2}.
\end{equation*}%
Since, the Kodama hydrodynamic type system is homogeneous, then the function 
$\lambda (u,\upsilon ,w,p)$ must be homogeneous. Then $\lambda =(2p\partial
_{p}+2u\partial _{u}+3\upsilon \partial _{\upsilon }+4w\partial _{w})\lambda 
$ up to some insufficient constant factor (degree of homogeneity). Thus, the
equation of the Riemann surface can be found in quadratures. For instance,%
\begin{equation*}
\partial _{p}\ln \lambda =\frac{2\Delta }{%
(p-u)[2p^{3}-4up^{2}+2(u^{2}+w)p-2uw+\upsilon ^{2}]}.
\end{equation*}%
Then (cf. \textbf{\cite{Kod+water}}) the equation of the Riemann surface for
the Kodama hydrodynamic type system is%
\begin{equation*}
\lambda =p+\frac{w}{p-u}+\frac{\upsilon ^{2}}{2(p-u)^{2}}.
\end{equation*}

\textbf{Remark}: The same procedure can be repeated for any $N$ component
Kodama hydrodynamic type system. Moreover, suppose the determinant (\textbf{%
\ref{8}}) is computed and written in the factorized form%
\begin{equation*}
\underset{k=1}{\overset{N}{\prod }}(p-p^{k}(\mathbf{a}))
\end{equation*}%
for an arbitrary $N$, then the equation of the Riemann surface is given by
rational function (see \textbf{\cite{Kod+water}})%
\begin{equation}
\lambda =\int \underset{k=1}{\overset{N}{\prod }}\frac{p-p^{k}(\mathbf{a})}{%
p-a^{1}}dp\equiv p+\underset{k=1}{\overset{N-1}{\sum }}\frac{B_{k}(\mathbf{a}%
)}{(p-a^{1})^{k}},  \label{q}
\end{equation}%
where $B_{k}$ are some polynomials with respect to flat coordinates $a^{n}$.
These coefficients $B_{k}$ can be found by substitution the above formula in
(\textbf{\ref{7}}). The corresponding linear system is%
\begin{equation*}
\partial _{n}B^{k+1}=\underset{m=n+1}{\overset{N}{\sum }}a^{m+1-n}\partial
_{m}B^{k}\text{, \ \ \ }k=1,2,...,N-2\text{, \ \ \ \ }n=2,3,...,N-1,
\end{equation*}%
where $B^{1}\equiv a^{N}$ and%
\begin{equation*}
\underset{m=n+1}{\overset{N-1}{\sum }}a^{m+1-n}\partial _{m}B^{N-1}=0\text{,
\ \ \ \ \ }\partial _{N}B^{n}=0\text{, \ \ \ \ }n=2,3,...,N-1.
\end{equation*}

\textbf{Remark}: In the symmetric case (\textbf{\ref{1}}) all derivatives $%
\lambda _{k}=(...)\lambda _{p}$ can be found immediately, but in the above
case for each $N$, one must (step by step) compute all above derivatives
consequently. If the given hydrodynamic type system is non-symmetric, if the
generating function of conservation laws (as in the above example) is not
given a priori, then derivation of the equation of the Riemann surface
becomes the very complicated computational problem.

\textbf{Remark}: The equation of the Riemann surface (\textbf{\ref{q}}) can
be written in the totally factorized form%
\begin{equation*}
\lambda =(p-a^{1})^{1-N}\underset{k=1}{\overset{N}{\prod }}(p-b^{k}),
\end{equation*}%
where $(N-1)a^{1}=\Sigma b^{k}(\mathbf{a})$. Substituting the Taylor series (%
\textbf{\ref{zak}}) $p^{(k)}=b^{k}+\lambda c^{k}(\mathbf{b})+...$ in (%
\textbf{\ref{con}}) yields the Kodama hydrodynamic type system written in
the symmetric form%
\begin{equation}
b_{t}^{k}=\partial _{x}\left[ \frac{(b^{k})^{2}}{2}+\frac{1}{2}\sum
(b^{k})^{2}-\frac{1}{2(N-1)}\left( \sum b^{k}\right) ^{2}\right] .
\label{tuk}
\end{equation}

\subsection{Cubic Hamiltonian hydrodynamic type system}

The \textbf{cubic Hamiltonian} hydrodynamic type system (\textbf{\ref{ham}})
with the Hamiltonian density%
\begin{equation*}
\mathbf{h}=\frac{1}{6}\sum \gamma _{k}(u^{k})^{3}+\beta \sum \gamma
_{k}u^{k}\sum \gamma _{n}(u^{n})^{2}+\varepsilon \left( \sum \gamma
_{k}u^{k}\right) ^{3}
\end{equation*}%
is equivalent to homogeneous hydrodynamic type system (cf. (\textbf{\ref{tuk}%
}))%
\begin{equation}
a_{t}^{i}=\partial _{x}\left[ \frac{(a^{i})^{2}}{2}+\alpha \sum \gamma
_{k}(a^{k})^{2}+\delta \left( \sum \gamma _{k}a^{k}\right) ^{2}\right]
\label{be}
\end{equation}%
under the transformation $a^{i}=u^{i}+2\beta \Sigma \gamma _{k}u^{k}$.

Following the recipe given above, one can verify that the hydrodynamic type
system (\textbf{\ref{be}}) is integrable iff%
\begin{equation*}
\delta =-\frac{2\alpha ^{2}}{1+2\alpha \sum \gamma _{n}}\text{ \ \ \ \ \ }%
\Longleftrightarrow \text{ \ \ \ \ \ }\varepsilon =-\frac{2\beta ^{2}/3}{%
1+2\beta \sum \gamma _{n}}
\end{equation*}%
and is connected with the equation of the Riemann surface%
\begin{equation*}
\lambda =\left( p-\frac{2\alpha }{1+2\alpha \sum \gamma _{n}}\sum \gamma
_{m}a^{m}\right) ^{1+2\alpha \sum \gamma _{s}}\prod (p-a^{k})^{-2\alpha
\gamma _{k}}.
\end{equation*}

\textbf{Remark}: \textit{The mechanical interpretation}. Let us consider the
classical Hamilton's system%
\begin{equation*}
\dot{x}=\frac{\partial \mathbf{H}}{\partial p}\text{, \ \ \ \ \ }\dot{p}=-%
\frac{\partial \mathbf{H}}{\partial x},
\end{equation*}%
where the Hamiltonian is $\mathbf{H}=p^{2}/2+V(x,t)$. We seek the
Hamiltonian system $\ddot{x}=-V_{x}$ possessing the first integral (``energy
constant surface''; see \textbf{\cite{Koz}})%
\begin{equation}
\lambda =V_{0}+V_{1}\dot{x}+V_{2}\dot{x}^{2}+...+V_{N-2}\dot{x}^{N-2}+V_{N}%
\dot{x}^{N},  \label{g}
\end{equation}%
where $V_{k}(x,t)$ are some functions and $N$ is arbitrary. Differentiating
this equation with respect to $t$ yields the hydrodynamic type system (%
\textbf{\ref{be}}) (cf. (\textbf{\ref{tuk}}))%
\begin{equation*}
u_{t}^{k}=-\partial _{x}\left( \frac{(u^{k})^{2}}{2}-\frac{1}{2(N+1)}\left[
\left( \sum u^{m}\right) ^{2}+\sum (u^{m})^{2}\right] \right) ,
\end{equation*}%
where the field variables $u^{k}$ are coefficients of the polynomial (%
\textbf{\ref{g}}) written in the factorized form%
\begin{equation*}
\lambda =(\dot{x}+\sum u^{m})\prod (\dot{x}-u^{k})
\end{equation*}%
and the ``potential energy'' is given by the symmetric expression%
\begin{equation*}
V=-\frac{1}{2(N+1)}\left[ \left( \sum u^{m}\right) ^{2}+\sum (u^{m})^{2}%
\right] .
\end{equation*}

Let us replace (\textbf{\ref{g}}) on an arbitrary dependence%
\begin{equation*}
\lambda =\lambda (\mathbf{u};\dot{x}),
\end{equation*}%
where $u^{k}(x,t)$ are some functions. Differentiating of the above equation
with respect to $t$, one can obtain (see (\textbf{\ref{0}}), (\textbf{\ref{8}%
}); cf. (\textbf{\ref{7}}))%
\begin{equation*}
\lambda _{\mu }\partial _{i}V(\mathbf{u})=(\upsilon _{i}^{k}(\mathbf{u})-\mu
\delta _{i}^{k})\lambda _{k},
\end{equation*}%
where we use the equation of the Riemann surface $\lambda =\lambda (\mathbf{u%
};\mu )$. In the symmetric case (\textbf{\ref{1}}) the above system reduces
to (see (\textbf{\ref{4}}), (\textbf{\ref{6}}); cf. (\textbf{\ref{7}}))%
\begin{equation*}
\lambda _{\mu }\partial _{i}V(\mathbf{u})=A_{i}^{k}(\mathbf{u},\mu )\lambda
_{k},
\end{equation*}%
which \textit{should be} connected with (\textbf{\ref{7}}) by the
substitution (see (\textbf{\ref{4}}))%
\begin{equation*}
\mu =\frac{\partial \psi }{\partial p}.
\end{equation*}

\subsection{The ideal gas dynamics}

The ideal gas dynamics%
\begin{equation}
u_{t}=\partial _{x}\left( \frac{u^{2}}{2}+\frac{\upsilon ^{\beta }}{\beta }%
\right) \text{, \ \ \ \ \ \ }\upsilon _{t}=\partial _{x}(u\upsilon )
\label{gas}
\end{equation}%
is connected with the equation of the Riemann surface (see \textbf{\cite{Das}%
})%
\begin{equation}
\lambda =\frac{1}{4}[p^{\beta }+\beta u+\upsilon ^{\beta }p^{-\beta }].
\label{sur}
\end{equation}%
The corresponding Gibbons equation is%
\begin{equation}
\lambda _{t}-(p^{\beta }+u)\lambda _{x}=\frac{\partial \lambda }{\partial p}%
\left[ p_{t}-\partial _{x}\left( \frac{p^{\beta +1}}{\beta +1}+up\right) %
\right] .  \label{gaz}
\end{equation}

\textbf{Remark}: The equation of the Riemann surface (\textbf{\ref{sur}}) is
unique for all values of index $\beta $. Indeed, introducing the Riemann
invariants%
\begin{equation*}
r^{1,2}=u\pm \frac{2}{\beta }\upsilon ^{\beta /2},
\end{equation*}%
the ideal gas dynamics (\textbf{\ref{gas}}) can be written in the diagonal
form%
\begin{equation*}
r_{t}^{1,2}=\frac{1}{2}[r^{1}+r^{2}\pm \frac{\beta }{2}%
(r^{1}-r^{2})]r_{x}^{1,2}.
\end{equation*}%
Then the equation of the Riemann surface (\textbf{\ref{sur}}) in the Riemann
invariants is%
\begin{equation}
\lambda =\frac{\nu }{4}+\frac{r^{1}+r^{2}}{2}+\frac{(r^{1}-r^{2})^{2}}{4\nu }%
,  \label{lya}
\end{equation}%
where%
\begin{equation*}
\nu =p^{\beta }.
\end{equation*}

\textbf{Remark}: Two infinite series of conservation laws can be obtained
with the aid of the B\"{u}rmann--Lagrange series (see the next section) at
the vicinity of two zeros of the equation of the Riemann surface%
\begin{equation*}
4\lambda =\frac{[\nu +(\sqrt{r^{1}}+\sqrt{r^{2}})^{2}][\nu +(\sqrt{r^{1}}-%
\sqrt{r^{2}})^{2}]}{\nu },
\end{equation*}%
while the Kruskal series can be derived at the infinity ($\lambda
\rightarrow \infty ,\nu \rightarrow \infty $) and at the vicinity of another
singular point ($\lambda \rightarrow \infty ,\nu \rightarrow 0$).

\subsection{Whitham averaged Sinh-Gordon equation}

A one-phase solution of the Sinh--Gordon%
\begin{equation*}
u_{xt}=\sinh u
\end{equation*}%
averaged by the Whitham approach is the two component hydrodynamic type
system (see \textbf{\cite{Whitham}})%
\begin{equation}
r_{t}^{i}=\mu ^{i}(\mathbf{r})r_{x}^{i}\text{,}  \label{sinh}
\end{equation}%
where the differentials of the quasi-momentum and the quasi-energy,
respectively,%
\begin{equation*}
dp=\frac{\lambda -<\lambda >}{\sqrt{\lambda (r^{1}-\lambda )(r^{2}-\lambda )}%
}d\lambda \text{, \ \ \ \ }dq=\frac{\frac{1}{\lambda }-<\frac{1}{\lambda }>}{%
\sqrt{\lambda (r^{1}-\lambda )(r^{2}-\lambda )}}d\lambda
\end{equation*}%
determine characteristic velocities%
\begin{equation*}
\mu ^{1,2}(\mathbf{r})=\frac{dq(\lambda )}{dp(\lambda )}|_{\lambda =r^{1,2}}=%
\frac{1}{\sqrt{r^{1}r^{2}}}\left[ 1-(1-s^{2})\frac{\mathbf{K}(s)}{\mathbf{E}%
(s)}\right] ^{\pm 1},
\end{equation*}%
where $\mathbf{K}(s)$ and $\mathbf{E}(s)$ are complete elliptic integrals of
the first and second kind, respectively; $s^{2}=r^{2}/r^{1}$ is elliptic
module, and%
\begin{equation*}
<a>\equiv \frac{1}{\mathbf{T}}\overset{r^{2}}{\underset{0}{\int }}\frac{%
ad\lambda }{\sqrt{\lambda (r^{1}-\lambda )(r^{2}-\lambda )}}\text{, \ \ \ \
\ \ }\mathbf{T=}\overset{r^{2}}{\underset{0}{\int }}\frac{d\lambda }{\sqrt{%
\lambda (r^{1}-\lambda )(r^{2}-\lambda )}}.
\end{equation*}

However, this hydrodynamic type system can be enclosed in the framework
presented in the previous sub-section. Two next theorems can be proved by
straightforward calculation.

\textbf{Theorem 4}: \textit{The Gibbons equation}%
\begin{equation*}
\lambda _{t}-\frac{d\tilde{q}/d\nu }{d\tilde{p}/d\nu }\lambda _{x}=\frac{%
\partial \lambda }{\partial \tilde{p}}(\tilde{p}_{t}-\tilde{q}_{x})
\end{equation*}%
\textit{connects the averaged (by the Whitham approach) one-phase solution
of the Sinh-Gordon equation }(\textbf{\ref{sinh}}) \textit{with the equation
of the Riemann surface} (\textbf{\ref{lya}}), \textit{where}%
\begin{equation*}
\tilde{p}=\frac{\nu }{2}+r^{1}+r^{2}+\left[ \frac{r^{1}+r^{2}}{2}-<\lambda >%
\right] \ln \nu \text{, \ \ \ \ \ }\tilde{q}=\ln \frac{\nu +(\sqrt{r^{1}}-%
\sqrt{r^{2}})^{2}}{\nu +(\sqrt{r^{1}}+\sqrt{r^{2}})^{2}}-\frac{<\lambda >}{%
\sqrt{r^{1}r^{2}}}\ln \nu ,
\end{equation*}%
where%
\begin{equation*}
<\lambda >=r^{1}\left[ 1-\frac{\mathbf{K}(s)}{\mathbf{E}(s)}\right] .
\end{equation*}

\textbf{Theorem 5}: \textit{The generating function of commuting flows in
the Riemann invariants}%
\begin{equation*}
r_{\tau (\zeta )}^{i}=\left( \ln \nu (\zeta )+g^{ii}\frac{\nu ^{i}+2P}{\nu
^{i}-\nu (\zeta )}\right) r_{x}^{i}
\end{equation*}%
\textit{is connected with the Gibbons equation}%
\begin{equation*}
\lambda _{\tau (\zeta )}-\left( \ln \nu (\zeta )+\frac{3\nu ^{2}(\lambda
)+4(r^{1}+r^{2})\nu (\lambda )+(r^{1}-r^{2})}{(\nu (\lambda )+2P)(\nu
(\lambda )-\nu (\zeta ))}\right) \lambda _{x}=\frac{\partial \lambda }{%
\partial \tilde{p}(\lambda )}\left[ \partial _{\tau (\zeta )}\tilde{p}%
(\lambda )-\partial _{x}Q(\lambda ,\zeta )\right] ,
\end{equation*}%
\textit{where the potential }$\mathbf{P}$\textit{\ of the Egorov metric} $%
g_{ii}=\partial _{i}\mathbf{P}$ \textit{is}%
\begin{equation*}
\mathbf{P}=\frac{r^{1}+r^{2}}{2}-<\lambda >,
\end{equation*}%
\textit{the Egorov metric is}%
\begin{equation*}
g_{11}=\frac{\mathbf{E}^{2}(s)/\mathbf{K}^{2}(s)}{2(1-s^{2})}\text{, \ \ \ \
\ \ \ }g_{22}=-\frac{[1-s^{2}-\mathbf{E}(s)/\mathbf{K}(s)]^{2}}{%
2s^{2}(1-s^{2})}
\end{equation*}%
\textit{and}%
\begin{eqnarray*}
Q(\lambda ,\zeta ) &=&2(r^{1}+r^{2})+\frac{1}{2}\nu (\lambda )+\frac{1}{2}%
\nu (\zeta )+\mathbf{P}\ln \nu (\lambda )\ln \nu (\zeta )+[r^{1}+r^{2}+\frac{%
1}{2}\nu (\zeta )]\ln \nu (\lambda ) \\
&& \\
&&+[r^{1}+r^{2}+\frac{1}{2}\nu (\lambda )]\ln \nu (\zeta )+2(\lambda +\zeta
)\ln [\nu (\lambda )-\nu (\zeta )]-2\lambda \ln \nu (\zeta )-2\zeta \ln \nu
(\lambda ).
\end{eqnarray*}

\section{Integrable hydrodynamic chains}

The Gibbons equation for the hydrodynamic type system (\textbf{\ref{poli}})%
\begin{equation}
\lambda _{t^{1}}-(p^{1-\beta \varepsilon }+\tilde{a})\lambda _{x}=\frac{%
\partial \lambda }{\partial p}\left[ p_{t^{1}}-\partial _{x}\left( \frac{%
p^{2-\beta \varepsilon }}{2-\beta \varepsilon }+\tilde{a}p\right) \right] ,
\label{gib}
\end{equation}%
where $\tilde{a}=a/(\beta -\beta \varepsilon +1)$ (see (\textbf{\ref{den}}%
)), is exactly the same as the Gibbons equation for the ideal gas dynamics (%
\textbf{\ref{gaz}}).

Introducing the moments%
\begin{equation*}
B^{k}=\frac{1}{(1-\beta \varepsilon )(k+1)+\beta }\sum \gamma
_{i}(u^{i})^{(1-\beta \varepsilon )(k+1)+\beta }\text{, \ \ \ \ \ }%
k=0,1,2...,
\end{equation*}%
the hydrodynamic type system (\textbf{\ref{poli}}) can be rewritten as the
Kupershmidt hydrodynamic chain (see \textbf{\cite{Kuper}})%
\begin{equation}
B_{t^{1}}^{k}=B_{x}^{k+1}+B^{0}B_{x}^{k}+[(1-\beta \varepsilon )(k+1)+\beta
]B^{k}B_{x}^{0}\text{, \ \ \ \ \ \ \ }k=0,1,2,...  \label{n}
\end{equation}%
connected with the Gibbons equation (\textbf{\ref{gib}}), where%
\begin{equation*}
\tilde{a}=\frac{B^{0}}{1-\beta \varepsilon }.
\end{equation*}

Suppose the moments $B^{k}$ of the Kupershmidt hydrodynamic chain are 
\textit{some} functions of $N$ field variables $u^{k}$, then $N$ component
hydrodynamic reduction%
\begin{equation}
u_{t^{1}}^{i}=\partial _{x}\left( \frac{(u^{i})^{2-\beta \varepsilon }}{%
2-\beta \varepsilon }+\frac{B^{0}(\mathbf{u})}{1-\beta \varepsilon }%
u^{i}\right)  \label{chrom}
\end{equation}%
is an \textit{integrable} hydrodynamic type system (\textbf{\ref{1}}) iff
the function $B^{0}(\mathbf{u})$ satisfies some nonlinear PDE system, which
is consequence of the compatibility condition $\partial _{i}(\partial
_{k}p)=\partial _{k}(\partial _{i}p)$, where (cf. (\textbf{\ref{comp}}))%
\begin{equation}
\partial _{i}p=p\frac{\partial _{i}B^{0}(\mathbf{u})}{(u^{i})^{1-\beta
\varepsilon }-p^{1-\beta \varepsilon }}\left[ 1-\beta \varepsilon +\sum 
\frac{\partial _{k}B^{0}(\mathbf{u})}{(u^{k})^{1-\beta \varepsilon
}-p^{1-\beta \varepsilon }}\right] ^{-1}.  \label{der}
\end{equation}

If $B^{0}(\mathbf{u})=(1-\beta \varepsilon )\Sigma \gamma _{k}(u^{k})^{\beta
-\beta \varepsilon +1}/(\beta -\beta \varepsilon +1)$, then (\textbf{\ref%
{der}}) reduces to (\textbf{\ref{comp}}). The compatibility condition $%
\partial _{i}(\partial _{k}p)=\partial _{k}(\partial _{i}p)$ creates a
nonlinear PDE system on function $B^{0}(\mathbf{u})$ only. Its solution is
parameterized by $N$ arbitrary functions of a single variable. Thus, any
symmetric hydrodynamic type system (\textbf{\ref{1}}) can be used for
derivation of corresponding integrable hydrodynamic chain, which have the
same Gibbons equation. At the same time, all hydrodynamic reductions can be
written in a similar symmetric form, but with another dependence of r.h.s.
functions like $B^{0}(\mathbf{u})$ with respect to field variables $u^{k}$.
If somebody will be able to solve corresponding nonlinear PDE system (which
is known in the Riemann invariants as the Gibbons--Tsarev system, see 
\textbf{\cite{Gib+Tsar}}), then infinitely many symmetric hydrodynamic type
systems (\textbf{\ref{1}}) will be produced.

\textbf{Corollary}: Let us consider the generating function of conservation
laws (see (\textbf{\ref{chrom}}))%
\begin{equation*}
p_{t^{1}}=\partial _{x}\left[ \frac{p^{\alpha +1}}{\alpha +1}+\frac{B^{0}(%
\mathbf{u})}{\alpha }p\right]
\end{equation*}%
and introduce the parameter $\alpha =1-\beta \varepsilon $. Then $N+1$
parametric family ($N$ parameters $\gamma _{k}$ and $\beta $ for each 
\textit{fixed} index $\alpha $) of hydrodynamic reductions of the
hydrodynamic chain (\textbf{\ref{n}})%
\begin{equation*}
B_{t^{1}}^{k}=B_{x}^{k+1}+B^{0}B_{x}^{k}+[\alpha (k+1)+\beta ]B^{k}B_{x}^{0}%
\text{, \ \ \ \ \ \ \ }k=0,1,2,...
\end{equation*}%
is a set of the hydrodynamic type systems (\textbf{\ref{chrom}})%
\begin{equation*}
u_{t^{1}}^{i}=\partial _{x}\left( \frac{(u^{i})^{\alpha +1}}{\alpha +1}+%
\frac{\sum \gamma _{k}(u^{k})^{\alpha +\beta }}{\alpha +\beta }u^{i}\right) ,
\end{equation*}%
which are distinct for every value of index $\beta $ (all above hydrodynamic
chains are equivalent for each fixed index $\alpha $ and for any value of
the index $\beta $, see details in \textbf{\cite{Maks+Kuper}}).

\textbf{Remark}: Introducing the moments%
\begin{equation*}
C^{k}=\frac{1}{(\beta \varepsilon -1)k+\beta }\sum \gamma
_{i}(u^{i})^{(\beta \varepsilon -1)k+\beta }\text{, \ \ \ \ \ }k=0,1,2...,
\end{equation*}%
the hydrodynamic type system (\textbf{\ref{ful}}) can be rewritten as the
Kupershmidt hydrodynamic chain (see \textbf{\cite{Kuper1}})%
\begin{equation}
C_{t}^{k}=\beta \varepsilon (1+\beta C^{0})^{-\varepsilon
}C_{x}^{k+1}+[(\beta \varepsilon -1)(k+1)+\beta ]C^{k+1}[(1+\beta
C^{0})^{-\varepsilon }]_{x}\text{, \ \ \ }k=0,1,2,...  \label{m}
\end{equation}

In this section we proved that hydrodynamic type systems (\textbf{\ref{ful}}%
) and (\textbf{\ref{poli}}) are commuting flows, then the corresponding ($B$
and $C$) hydrodynamic chains are commute with each other. All other details
can be found in \textbf{\cite{Maks+Kuper}}.

\section{Hamiltonian chromatography system}

Another generalization of the chromatography system (cf. (\textbf{\ref{hrom}}%
) and \textbf{\cite{Fer+Tsar}}) is given by the Hamiltonian hydrodynamic
type system%
\begin{equation*}
a_{t}^{i}=\partial _{x}\frac{\partial \mathbf{h}}{\partial a^{i}}\text{, \ \
\ \ \ }i=1,2,...,N,
\end{equation*}%
where the Hamiltonian density $\mathbf{h}(\Delta )$ and $\Delta =\Sigma
z_{k}(a^{k})$. If this system is diagonalizable (see (\textbf{\ref{ri}}))%
\begin{equation*}
r_{t}^{i}=\mu ^{i}(\mathbf{r})r_{x}^{i}\text{, \ \ \ \ \ \ }i=1,2,...,N,
\end{equation*}%
then it is integrable (see \textbf{\cite{Dubr+Nov}}). Thus, we are looking
for corresponding transformation $r^{i}(\mathbf{a})$. A direct computation
yields (the indexes of the Riemann invariants $r^{k}$ and characteristic
velocities $\mu ^{i}$ are omitted for simplicity below)%
\begin{equation*}
\frac{\partial r}{\partial z_{i}}=\frac{\varphi }{\zeta -V_{i}^{\prime }}%
\text{, \ \ \ \ \ \ }\rho (\Delta )=\sum \frac{V_{n}}{\zeta -V_{n}^{\prime }}%
,
\end{equation*}%
where $V_{k}(z_{k})=z_{k}^{\prime ^{2}}/2$, $(\ln \mathbf{h}^{\prime
})^{\prime }\rho (\Delta )=1/2$, $\varphi =\rho ^{-1}(\Delta )\Sigma
V_{k}\partial r/\partial z_{k}$ and $\mu =\zeta \mathbf{h}^{\prime }$. The
Riemann invariants exist iff the compatibility conditions $\partial
_{i}(\partial _{k}r)=\partial _{k}(\partial _{i}r)$ are fulfilled, where $%
\partial _{i}\equiv \partial /\partial z_{k}$. Eliminating $\varphi $ and
its first derivatives from the compatibility conditions, one can obtain the 
\textit{integrability condition}%
\begin{equation*}
q_{j}(\partial _{i}q_{k}-\partial _{k}q_{i})+q_{k}(\partial
_{j}q_{i}-\partial _{i}q_{j})+q_{i}(\partial _{k}q_{j}-\partial _{j}q_{k})=0
\end{equation*}%
for every three distinct indexes $i,j,k$, where $q_{i}=(\zeta -V_{i}^{\prime
})^{-1}$. Taking into account%
\begin{equation*}
\partial _{i}\zeta =\left( \frac{V_{i}V_{i}^{\prime \prime }}{(\zeta
-V_{i}^{\prime })^{2}}+\frac{V_{i}^{\prime }}{\zeta -V_{i}^{\prime }}-\rho
^{\prime }(\Delta )\right) \left[ \sum \frac{V_{m}}{(\zeta -V_{m}^{\prime
})^{2}}\right] ^{-1},
\end{equation*}%
the above integrability condition reduces to sole ODE%
\begin{equation}
VV^{\prime \prime }=(1+\alpha )V^{\prime ^{2}}+\beta V^{\prime }+\gamma ,
\label{ode}
\end{equation}%
where $V_{i}\equiv V(z_{i})$, \ $z_{i}\equiv z(a^{i})$ and $\rho (\Delta
)=\alpha \Delta +\delta $ ($\alpha ,\beta ,\gamma ,\delta $ are arbitrary
constants). Thus, the Hamiltonian chromatography system%
\begin{equation}
a_{t}^{i}=\partial _{x}[\mathbf{h}^{\prime }(\Delta )z^{\prime }(a^{i})]
\label{gam}
\end{equation}%
is integrable if $\mathbf{h}=e^{\Delta }$ ($\alpha =0$), $\mathbf{h}=\ln
\Delta $ ($\alpha =-1/2$) and $\mathbf{h}=\Delta ^{\varepsilon }$ (for all
other values $\alpha $; $\varepsilon $ is an arbitrary constant).

Then the Gibbons equation%
\begin{equation*}
\lambda _{t}-z^{\prime \prime }(p)\mathbf{h}^{\prime }(\Delta )\lambda _{x}=%
\frac{\partial \lambda }{\partial p}\left( p_{t}-\partial _{x}[z^{\prime }(p)%
\mathbf{h}^{\prime }(\Delta )]\right)
\end{equation*}%
is determined by the equation of the Riemann surface $\lambda (\mathbf{a};p)$%
, which can be found in quadratures%
\begin{equation*}
d\lambda =z^{1+4\alpha }(p)\exp [\beta \int \frac{dz}{V}]\left[ z^{\prime
}(p)\sum \frac{dz_{n}}{z^{\prime \prime }(p)-V_{n}^{\prime }}+2\left( \alpha
\Delta +\delta -\sum \frac{V_{n}}{z^{\prime \prime }(p)-V_{n}^{\prime }}%
\right) dp\right] ,
\end{equation*}%
where $\zeta =z^{\prime \prime }(p)$.

\textbf{Remark}: Introducing the moments%
\begin{equation*}
A^{k}=\sum \int V_{i}^{\prime ^{k}}dz_{i}
\end{equation*}%
the Hamiltonian chromatography system (\textbf{\ref{gam}}) can be rewritten
as the Hamiltonian hydrodynamic chain%
\begin{equation*}
A_{t}^{k}=\underset{n=0}{\overset{k+1}{\sum }}F_{n}^{k}(\mathbf{A})A_{x}^{n}%
\text{, \ \ \ \ \ \ }k=0,1,2,...
\end{equation*}%
determined by the Hamiltonian density $\mathbf{h}(A^{0})$ ($\mathbf{h}%
=e^{A^{0}}$, $\mathbf{h}=\ln A^{0}$, $\mathbf{h}=(A^{0})^{\varepsilon }$)
and by the Poisson bracket%
\begin{equation*}
\{A^{k},A^{n}\}=[B_{k,n}\partial _{x}+\partial _{x}B_{n,k}]\delta
(x-x^{\prime }),
\end{equation*}%
where (here we use (\textbf{\ref{ode}}))%
\begin{equation*}
B_{k,n}=[2(1+\alpha )k+1]A^{k+n+1}+2\beta kA^{k+n}+2\gamma kA^{k+n-1}.
\end{equation*}

\section{Exceptional (linearly degenerate) case}

In this paper the algebro-geometric approach for integrability of
hydrodynamic type systems is established. However, this approach is most
effective just in case of the symmetric hydrodynamic type systems possessing
any symmetry operator (see section \textbf{6}). Then the generating function
of conservation laws for the symmetric hydrodynamic type systems can be
found immediately; the integration factor in the computation of the Riemann
surface can be found just if a corresponding hydrodynamic type system is
invariant under some Lie group of symmetries (like homogeneity).

Nevertheless, this algebro-geometric approach can be used in all other more
general and complicated cases, \textit{except possibly hydrodynamic type
systems which are hydrodynamic reductions of \textbf{linear-degenerate}
hydrodynamic chains} (see \textbf{\cite{Fer+Kar}}, \textbf{\cite{Maks+eps}}%
). These hydrodynamic type systems usually can be written explicitly in the
Riemann invariants (see \textbf{\cite{Maks+eps}}) and the conservation law 
\textit{fluxes} of their generating functions of conservation laws are 
\textit{linear} functions with respect to conservation law density $p$ (see 
\textbf{\cite{Maks+eps}})%
\begin{equation}
p_{t}=\partial _{x}[p\upsilon (\mathbf{r};\lambda )].  \label{un}
\end{equation}

For instance, in the limit $\beta \varepsilon =1$, the integrable
chromatography system (\textbf{\ref{ful}})%
\begin{equation*}
u_{t}^{i}=\varepsilon _{i}\partial _{x}\frac{u^{i}}{[1+\sum \gamma
_{k}(u^{k})^{1/\varepsilon }]^{\varepsilon }}\text{, \ \ \ \ \ }i=1,2,...,N
\end{equation*}%
can be written explicitly in the Riemann invariants (see \textbf{\cite%
{Fer+Tsar}}, the formula \textbf{5}; also \textbf{\cite{Maks+eps}}; the
parameters $\varepsilon _{i}$ and $\gamma _{k}$ do not affect on explicit
expressions of characteristic velocities in the Riemann invariants)%
\begin{equation*}
r_{t}^{i}=\frac{\left( \prod r^{m}\right) ^{\varepsilon }}{r^{i}}r_{x}^{i}%
\text{, \ \ \ \ \ \ }i=1,2,...,N.
\end{equation*}%
Its generating function of conservation laws is (see \textbf{\cite{Maks+eps}}%
)%
\begin{equation*}
p_{t}=\partial _{x}\left( p\frac{\left( \prod r^{m}\right) ^{\varepsilon }}{%
\lambda }\right) .
\end{equation*}%
In the same limit $\beta \varepsilon =1$, the hydrodynamic type system (%
\textbf{\ref{poli}})%
\begin{equation*}
u_{t^{1}}^{i}=\delta _{i}\partial _{x}\left[ u^{i}\left( 1+\varepsilon \sum
\gamma _{n}(u^{n})^{1/\varepsilon }\right) \right]
\end{equation*}%
in the Riemann invariants is (the parameters $\delta _{i}$ and $\gamma _{k}$
do not affect on explicit expressions of characteristic velocities in the
Riemann invariants)%
\begin{equation*}
r_{t^{1}}^{i}=\left( r^{i}-\varepsilon \sum r^{m}\right) r_{x}^{i}.
\end{equation*}%
Its generating function of conservation laws is (see \textbf{\cite{Maks+eps}}%
)%
\begin{equation*}
p_{t^{1}}=\partial _{x}\left[ p\left( \lambda -\varepsilon \sum r^{m}\right) %
\right] .
\end{equation*}

In this section we consider hydrodynamic type system written \textit{%
explicitly} in the Riemann invariants%
\begin{equation*}
r_{t}^{i}=\upsilon ^{i}(\mathbf{r})r_{x}^{i},\ \ \ \ \ \ \ i=1,2,...,N,
\end{equation*}%
such that the characteristic velocities $\upsilon ^{i}(\mathbf{r})$ are
determined by the unique function $\upsilon (\mathbf{r;}\lambda )$%
\begin{equation}
\upsilon ^{i}(\mathbf{r})=\upsilon (\mathbf{r;}\lambda )|_{\lambda =r^{i}}.
\label{uni}
\end{equation}%
Then the semi-Hamiltonian criterion (\textbf{\ref{semi}}) reduces to%
\begin{equation}
\partial _{j}\frac{\partial _{i}\upsilon }{\upsilon _{i}-\upsilon }=\partial
_{i}\frac{\partial _{j}\upsilon }{\upsilon _{j}-\upsilon }\text{, \ \ \ \ \ }%
i\neq j.  \label{polu}
\end{equation}

Suppose this hydrodynamic type system has the generating function of
conservation laws (\textbf{\ref{un}}). Then%
\begin{equation*}
\partial _{i}\ln p=\frac{\partial _{i}\upsilon }{\upsilon _{i}-\upsilon }
\end{equation*}%
and the semi-Hamiltonian criterion (\textbf{\ref{polu}}) is satisfied
automatically. Moreover, the generating function of conservation law
densities $p$ in such case can be found explicitly%
\begin{equation*}
p=\exp \left[ \int \sum \frac{\partial _{k}\upsilon }{\upsilon _{k}-\upsilon 
}dr^{k}\right] .
\end{equation*}%
Thus, any commuting flow%
\begin{equation*}
r_{\tau }^{i}=w^{i}(\mathbf{r})r_{x}^{i}
\end{equation*}%
has similar generating function of conservation laws%
\begin{equation*}
p_{\tau }=\partial _{x}[pw(\mathbf{r,}\lambda )],
\end{equation*}%
where $w^{i}(\mathbf{r})=w(\mathbf{r,}\lambda )|_{\lambda =r^{i}}$.

\textbf{Remark}: $N$ phase solutions of the nonlinear equations (like KdV,
NLS, Sinh-Gordon) averaged by the Whitham approach are hydrodynamic type
systems belong to the class presented in this section (\textbf{\ref{uni}})%
\begin{equation*}
r_{t}^{i}=\frac{dq}{dp}|_{\lambda =r^{i}}r_{x}^{i},
\end{equation*}%
where the Abelian holomorphic differentials of the quasi-momentum $dp$ and
the quasi-energy $dq$ are determined on the Riemann surfaces of genus $N$
(see, for instance, details in \textbf{\cite{Krich+kp}}). Thus,
corresponding hydrodynamic chains should be linear degenerate.

\section{Conclusion and outlook}

In this paper we suggest \textit{universal} (except linearly degenerate
case) approach for integrable (semi-Hamiltonian) symmetric hydrodynamic type
systems. This approach is very effective because the generating function of
conservation laws in such case is \textbf{unique}, this is a consequence of
the construction ((\textbf{\ref{1}}) $\rightarrow $ (\textbf{\ref{2}})). In
all other cases the first step is a computation of the generating function
of conservation laws.

Let us illustrate a complexity of this problem on two component hydrodynamic
type system (for simplicity and without lost of generality we can restrict
our consideration on two component case only). Suppose we have nonlinear
elasticity equation (this is nothing but the ideal gas dynamic written in
the Lagrangian coordinates, while (\textbf{\ref{gas}}) is written in the
Euler coordinates), which is commuting flow to ideal gas dynamics 
\begin{equation}
\upsilon _{y}=u_{x}\text{, \ \ \ \ \ \ }u_{y}=\partial _{x}\frac{\upsilon
^{\beta -1}}{\beta -1}.  \label{zu}
\end{equation}%
This hydrodynamic type system is written in non-symmetric form. Thus, we do
not know in advance the generating function of conservation laws in this
case. Nevertheless, obviously, we must seek such generating function in the
form%
\begin{equation}
p_{y}=\partial _{x}\psi (u,\upsilon ,p).  \label{gena}
\end{equation}%
However, if such generating function will be found, then this generating
function is a generating function for \textbf{whole} hydrodynamic chain, is
not for nonlinear elasticity only. Thus, one should seek $N$ component
hydrodynamic type systems written in the Riemann invariants (\textbf{\ref%
{rim}}) compatible with the above generating function. Thus, we have%
\begin{equation*}
\partial _{i}p=\frac{\frac{\partial \psi }{\partial u}\partial _{i}u+\frac{%
\partial \psi }{\partial \upsilon }\partial _{i}\upsilon }{\frac{\partial
\psi }{\partial p}|_{p=p^{i}}-\frac{\partial \psi }{\partial p}}.
\end{equation*}%
Here is a \textbf{crucial point} of the general construction. From the first
equation of the nonlinear elasticity equation (\textbf{\ref{zu}}) one can
obtain (take into account (\textbf{\ref{rim}}))%
\begin{equation}
\frac{\partial \psi }{\partial p}|_{p=p^{i}}\partial _{i}\upsilon =\partial
_{i}u.  \label{e}
\end{equation}%
From the second equation one gets%
\begin{equation}
\frac{\partial \psi }{\partial p}|_{p=p^{i}}\partial _{i}u=\upsilon ^{\beta
-2}\partial _{i}\upsilon .  \label{j}
\end{equation}%
Thus, we have two choices%
\begin{equation*}
\partial _{i}p=\frac{\frac{\partial \psi }{\partial u}\frac{\partial \psi }{%
\partial p}|_{p=p^{i}}+\frac{\partial \psi }{\partial \upsilon }}{\frac{%
\partial \psi }{\partial p}|_{p=p^{i}}-\frac{\partial \psi }{\partial p}}%
\partial _{i}\upsilon \text{, \ \ \ \ \ \ }\partial _{i}p=\frac{\frac{%
\partial \psi }{\partial u}+\upsilon ^{2-\beta }\frac{\partial \psi }{%
\partial \upsilon }\frac{\partial \psi }{\partial p}|_{p=p^{i}}}{\frac{%
\partial \psi }{\partial p}|_{p=p^{i}}-\frac{\partial \psi }{\partial p}}%
\partial _{i}u.
\end{equation*}%
These two options are no longer equivalent (in comparison with the symmetric
case), because we \textbf{ignore} the identity (which is consequence of (%
\textbf{\ref{e}}) and (\textbf{\ref{j}}))%
\begin{equation*}
\left( \frac{\partial \psi }{\partial p}|_{p=p^{i}}\right) ^{2}=\upsilon
^{\beta -2},
\end{equation*}%
which is valid for ideal gas dynamics only. Since we consider the generating
function of conservation laws (\textbf{\ref{gena}}) for whole hydrodynamic
chain, then depending on \textit{this} choice, we shall be able to construct 
\textbf{two} different hydrodynamic chains. Thus, for instance, the
nonlinear elasticity equation (\textbf{\ref{zu}}) can be \textit{embedded}
into different Kupershmidt hydrodynamic chains (\textbf{\ref{n}}) and (%
\textbf{\ref{m}}).

\textbf{Example}: The dispersionless limit of the Boussinesq system (see (%
\textbf{\ref{zu}}), $\beta =3$; \textbf{\cite{Gib+Kod}}, \textbf{\cite%
{Maks+eps}})%
\begin{equation}
\upsilon _{y}=u_{x}\text{, \ \ \ \ \ }u_{y}=\partial _{x}(\upsilon ^{2}/2)
\label{ga}
\end{equation}%
satisfies the Gibbons equation%
\begin{equation*}
\lambda _{y}-\frac{\upsilon ^{2}}{p^{3}}\lambda _{x}=\frac{\partial \lambda 
}{\partial p}\left[ p_{y}+\partial _{x}\frac{\upsilon ^{2}}{2p^{2}}\right] ,
\end{equation*}%
where $\mu =p^{3}$ and the equation of the Riemann surface is (cf. (\textbf{%
\ref{sur}}))%
\begin{equation}
\lambda =\mu +3u+\frac{\upsilon ^{3}}{\mu }.  \label{ku}
\end{equation}%
Simultaneously, the simplest reduction of the Benney moment chain (see 
\textbf{\cite{Gib+Kod}}) is again the dispersionless limit of the Boussinesq
system determined by the Gibbons equation%
\begin{equation*}
\lambda _{y}-\tilde{\mu}\lambda _{x}=\frac{\partial \lambda }{\partial 
\tilde{\mu}}\left[ \tilde{\mu}_{y}-\partial _{x}\left( \frac{\tilde{\mu}^{2}%
}{2}-\upsilon \right) \right] ,
\end{equation*}%
where the equation of the Riemann surface is%
\begin{equation}
\lambda =-\tilde{\mu}^{3}+3\upsilon \tilde{\mu}+3u.  \label{bus}
\end{equation}%
The substitution $\tilde{\mu}=-(p+\upsilon /p)$ in this cubic equation
yields exactly (\textbf{\ref{ku}})%
\begin{equation*}
\lambda =p^{3}+3u+\frac{\upsilon ^{3}}{p^{3}}.
\end{equation*}%
Factorizing the cubic polynomial (\textbf{\ref{bus}})%
\begin{equation*}
\lambda =-\tilde{p}(\tilde{p}-u^{1})(\tilde{p}-u^{2}),
\end{equation*}%
where $\tilde{p}=\tilde{\mu}+(u^{1}+u^{2})/3$, the dispersionless limit of
the Boussinesq system (\textbf{\ref{ga}})%
\begin{equation*}
u_{y}^{1}=\frac{1}{6}\partial _{x}[(u^{1})^{2}-2u^{1}u^{2}]\text{, \ \ \ \ \
\ }u_{y}^{2}=\frac{1}{6}\partial _{x}[(u^{2})^{2}-2u^{1}u^{2}]
\end{equation*}%
satisfies the Gibbons equation%
\begin{equation*}
\lambda _{y}-\left( \tilde{p}-\frac{u^{1}+u^{2}}{3}\right) \lambda _{x}=%
\frac{\partial \lambda }{\partial \tilde{p}}\left[ \tilde{p}_{y}-\partial
_{x}\left( \frac{\tilde{p}^{2}}{2}-\frac{u^{1}+u^{2}}{3}\tilde{p}\right) %
\right] .
\end{equation*}

\textbf{Remark}: Generating functions of conservation laws%
\begin{equation*}
p_{y}=-\partial _{x}\frac{\upsilon ^{2}}{2p^{2}}\text{, \ \ \ \ \ \ \ }%
\tilde{p}_{y}=\partial _{x}\left( \frac{\tilde{p}^{2}}{2}-\frac{u^{1}+u^{2}}{%
3}\tilde{p}\right)
\end{equation*}%
have \textit{different} sets of conservation law densities $\mathbf{H}_{k}$,
which \textbf{coincide} for nonlinear elasticity equation (\textbf{\ref{ga}}%
) up to insufficient factors.

In another paper we present a classification of integrable hydrodynamic
chains based on the concept of generating functions of conservation laws.
For instance, all generating functions of conservation laws (\textbf{\ref%
{gena}}) can be found. Thus, at least two of them are connected with the
ideal gas dynamics (\textbf{\ref{gas}}); each two component hydrodynamic
type system (\textbf{\ref{0}}) must be connected with some function $\psi
(u,\upsilon ,p)$.

We cannot suggest the recipe how to construct this function $\psi
(u^{1},u^{2},...,u^{N};p)$ for any a priori given hydrodynamic type system (%
\textbf{\ref{0}}) in general case. However, for any given function $\psi
(u^{1},u^{2},...,u^{N};p)$ we are able to reconstruct a corresponding
hydrodynamic type system (\textbf{\ref{0}}) together with its commuting
flows.

\section*{Acknowledgement}

I thank Boris Dubrovin, Eugeni Ferapontov, John Gibbons, Yuji Kodama, Boris
Kupershmidt, Sergey Tsarev, Vladimir Zakharov and Mikhail Zhukov for their
stimulating and clarifying discussions.

I am grateful to the Institute of Mathematics in Taipei (Taiwan) where some
part of this work has been done, and especially to Jen-Hsu Chang, Jyh-Hao
Lee, Ming-Hien Tu and Derchyi Wu for fruitful discussions.

\end{document}